\documentclass{article}
\usepackage[left=2cm,right=2cm,top=2cm,bottom=2cm]{geometry}
\usepackage{float}
\usepackage{amsmath}
\usepackage[english]{babel}
\usepackage[utf8]{inputenc}
\usepackage{amsfonts}
\usepackage{graphicx} 
\usepackage[square,numbers]{natbib}
\usepackage{subcaption}
\usepackage{hyperref}
\usepackage{booktabs}
\usepackage{authblk}

\newcommand{\Rb}{\mathbb{R}}

\newcommand{\Lc}{\mathcal{L}}
\DeclareMathOperator*{\argmin}{arg\,min}

\title{Clustering Digital Assets Using Path Signatures:
Application to Portfolio Construction}
\author[1]{Hugo Inzirillo}

 \affil[1]{
   CREST, Institut Polytechnique de Paris}

\date{October 2024}

\begin{document}

\maketitle
\begin{abstract}
We propose a new way of building portfolios of cryptocurrencies that provide good diversification properties to investors. 
First, we seek to filter these digital assets by creating some clusters based on their path signature. The goal is to identify similar patterns in the behavior of these highly volatile assets. Once such clusters have been built, we propose ``optimal'' portfolios by comparing the performances of such portfolios to a universe of unfiltered digital assets. Our intuition is that clustering based on path signatures will make it easier to capture the main trends and features of a group of cryptocurrencies, and allow parsimonious portfolios that reduce excessive transaction fees. Empirically, our assumptions seem to be satisfied. 
\end{abstract}

\section{Introduction}

Optimal portfolio construction research has driven the asset management industry since decades. Many innovative approaches have been proposed. Among all these approaches, mean-variance portfolio and maximum diversification strategies stand out for their intuitive appeal and historical effectiveness. During the last decades, new asset class emerged, in particular digital assets. These new assets differ from other asset classes by their very high volatility and peculiar dynamics (sequences of booms and bursts, for instance). The high volatility of digital assets is due to several factors probably linked to their youth, including regulatory uncertainty, speculation and the low liquidity in these new markets. 
Several researchers (\cite{bukovina2016sentiment,baur2017realized,pichl2017volatility,inzirillo2022attention}, etc.) have been studying the behavior and modeling of latent processes to determine the inherent risks of crypto-assets. Other researchers (\cite{balcilar2017can}, e.g.) have tried to introduce covariates to predict the performance or the volatility of digital assets. More recently,~\cite{schnoering2022deep} studied synchronous and asynchronous relationships using lead-lag graphs to detect some unexplained dependencies between digital assets. Nonetheless, no consensus has emerged in terms as a reliable model for predicting the price of these new assets, particularly in a multivariate setting. Indeed, the factors that influence their prices dynamics are numerous and complicated to understand. Considering many of them simultaneously seems to be a very difficult task. 

\medskip

For these reasons,
we will not try to explain/model the price dynamics of crypto assets strictly speaking, for instance with factors and/or covariates as in financial econometrics.
We will rather focus on grouping the numerous existing crypto assets into homogeneous classes. 
By selecting a representative asset in each class, we build a ``parsimonious'' portfolio of crypto assets that could behave better than other more 
``naive'' or standard portfolios.
Indeed, correlation-based clustering has been used to deduce hierarchical connections among diverse asset classes by analyzing the correlation matrix of their returns (\cite{bonanno2001high,song2019cluster,bonanno2003topology,mantegna1999hierarchical}), with promising results. 
The contribution of this research lies in its innovative approach to portfolio construction, leveraging  path signatures to enhance traditional methodologies. We explore how clustering based on path signatures can refine the asset selection process for mean-variance \cite{markowitz2000mean} and maximum diversification \cite{choueifaty2008toward} portfolios, potentially leading to superior risk-adjusted returns. Here, We will create homogenous clusters on the basis of path signatures, enabling the grouping of digital assets according to their risks and return profiles. The summary of information provided by signatures will enable portfolio managers to process and manage relatively few assets (compared to a naive mean-variance approach, e.g.).
Through empirical analysis and computational experiments, we demonstrate the effectiveness of our methodology. 

\medskip

The first step is to estimate path signatures and to create ``agnostic'' features solely based on historical prices. Path signatures~\cite{chen1958integration} are sequences of numbers that describe curves, here some price sequences over fixed periods of time. They are directly calculated from observed prices. They capture and summarize some empirical characteristics as the slopes and shapes of price curves, in addition to dependencies between successive quotes. The second step involves defining clusters based on these signatures to create an universe of investable digital assets, where each cluster represents a group of assets with ``similar'' path signatures. Clustering techniques aim to group similar objects according to their characteristics. Such procedures are widely used in various fields and constitute a classical research domain in statistics and machine learning. Here, the objective of clustering is to identify similar price patterns, structures, or relationships among sets of crypto currencies. The different steps of the methodology will be detailed in the next sections. Note that the idea of applying clustering in finance to build  optimized portfolios, i.e. to find the right trade-off between diversification and the cost of dynamically managing a large number of assets, is not new: see~\cite{leon2017clustering,korzeniewski2018efficient} or~\cite{marvin2015creating}. 

\medskip 

Even if the goal is to manage a reduced number of crypto assets in a portfolio, we still hope to benefit from some diversification effects (inside the 
univers of digital assets). Indeed, when it deals with portfolio diversification many advantages come to our mind:

\medskip

{\it Risk reduction}: this is the main argument in favor of diversification. By spreading investments across different assets that are not perfectly correlated, we can reduce the overall risk of the portfolio. If one investment performs poorly, the others can compensate. Digital assets, because of their youth and their nature, behave very different from more standard assets. Generally speaking, it is fruitful to further diversify portfolios by integrating assets whose behavior is as far as possible unrelated to the market.

\medskip

{\it Return potential}: by adding numerous assets into a portfolio, there a hope of including specific assets that are likely to outperform the others at a given time.

\medskip

{\it Access to broad investment opportunities}: a diversified portfolio typically includes equities, bonds, real estate assets, commodities, etc., providing access to a wide range of investment opportunities (\cite{rugman1976risk}). Despite the relatively low correlation of the digital asset market with traditional markets, its excess volatility is tarnishing its ability to reduce risk, and is even becoming a performance driver~\cite{baur2022bitcoin}. Path signatures give us a different representation of these time series to identify the relationships or common behaviors between digital assets. 

\medskip

\section{Path Signatures}

Introduced by~\cite{chen1958integration}, path signatures are sequences of iterated integrals of (transformed) time series (here, series of digital asset prices).
For a detailed presentation of path signatures as a reliable representation or a set of characteristics for unparameterized paths, we refer the reader to \cite{lyons2014rough,lyons1998differential,chevyrev2016primer}. A path signature, sometimes simply called a signature, is actually a mathematical representation that ``succinctly'' summarizes the pattern of a path. Although derived from rough path theory and stochastic analysis, it has been recently adopted in many other fields, as in machine learning (\cite{chevyrev2016primer,perez2018signature,fermanian2021embedding}), time series analysis (\cite{gyurko2013extracting,dyer2021deep}), computer vision (\cite{yang2022developing,li2017lpsnet}), etc.

\subsubsection{Definition}
\label{def_signatures}
 Let us define a N-dimensional path $(X_t)_{t \in [ 0,T ]}$. In other words, $X_t= (X_{t}^{1},...,X_{t}^{N})$. The  1-dimensional coordinate paths will be denoted as $ (X^n_t)$, $n\in \{1,\ldots,N\}$.
 To iteratively build the path signature of $(X_t)$, we first consider the increments of $X^1, \ldots, X^N$ over any interval $[0, t]$, $t \in [0, T]$, which are denoted $S(X)^1_{0,  t}, \ldots, S(X)^N_{0, t}$ and defined as follows:
       \begin{equation}
           S(X)^n_{0, t} :=  \int_0^{t} dX^n_s.
       \end{equation}
       
$S(X)^n_{0, t}$ is the first stage to calculate the signature of an unidimensional path. It is a sequence of real numbers, each of these numbers corresponding to an iterated integral of the path. 
Then, the next signature coefficients will involve two paths: a coordinate path $(X^m_t)$ and the ``increment path{{}} $(S(X)^n_{0, t})$ associated to the coordinate path $(X_t^n)$. 
There are $N^2$ such second order integrals, which are denoted $S(X)^{1,1}_{0, t} \ldots, S(X)^{N,N}_{0, t}$, where

\begin{equation}
    S(X)^{n,m}_{0, t} := \int_0^t S(X)^n_{0, s} \, dX^m_s,\; n,m\in\{1,\ldots,N\}. 
\end{equation}

\subsubsection{$k$-level path signatures}

 The set of first (resp. second) order integrals involves $N$ (resp. $N^2$) integrals. The latter first and second order integrals are called the first and second levels of path signatures respectively. Iteratively, we obtain 
 $N^k$ integrals of order $k$, which is denoted $S(X)^{i_{1},..,i_{k}}_{0, t}$ for the $k$-th level of path signatures, when $i_j\in \{1,\ldots,N\}$ and $j\in \{1,\ldots, k\}$.
       More precisely, the $k$-th level signatures can be written as follows: 
       $$
       S(X)^{i_{1},\ldots,i_{k}}_{0, t} :=  \int_0^t S(X)^{i_1, \ldots, i_{k-1}}_{0, s} \, dX^{i_k}_s.
       $$
    
    The path signature $S(X)_{0,T}$ is finally the infinite ordered set of such terms when considering all levels $k\geq 1$ and the path on the whole interval $[0,T]$: 
    \begin{equation}
    \begin{split}
            S(X)_{0,T} & := (1, S(X)^1_{0,T}, S(X)^2_{0,T}, \ldots, S(X)^N_{0,T}, S(X)^{1, 1}_{0,T},\\
            & S(X)^{1, 2}_{0,T}, \ldots, S(X)^{N, N}_{0,T}, S(X)^{1, 1, 1}_{0,T}, \ldots).
    \end{split}
    \end{equation}

\subsection{Properties of path signatures}

\subsubsection*{Uniqueness}

Obviously, there exists a single signature path associated to $(X_t)_{t\in [0,T]}$.
The uniqueness of a path signature lies in its ability to compactly and efficiently encode path's information. It means that a path signature can exhaustively capture all ``features'' of a path, including its geometry, the directions it tends to follow, the ``speed'' at which it moves, etc. 
The uniqueness property of signatures is explored in greater detail in~\cite{hambly2010uniqueness,boedihardjo2016signature}.

\subsubsection*{Translation invariance}

The value of the path integral $\int_a^b Y_t \, dX_t$ is invariant to $X$-translation. In other words, it does not change if we shift the entire path $(X_t)$:
\begin{equation}
    \int_a^b Y_t \, dX_t = \int_a^b Y_t \, dZ_t,
\end{equation}
where $Z_t = X_t + c$ for every $t$, for some constant $c$.

\subsubsection*{Reparametrisation invariance}

Let us define  $\psi: [a, b] \to [a, b]$ a function that is onto, continuous, and monotonically non-decreasing. The latter map is termed a ``reparametrization''. Let us consider two paths $X, Y: [a, b] \to \Rb$, and let $\psi: [a, b] \to [a, b]$ be a reparametrization of these paths. We can construct new paths $\tilde{X}, \tilde{Y}$: $[a, b] \to \Rb$, defined by the expressions $\tilde{X}_t = X_{\psi(t)}$ and $\tilde{Y}_t = Y_{\psi(t)}$ for any $t$. It is noted that

\begin{equation}
    d\tilde{X}_t =  \dot{\psi}(t) \,d X_{\psi(t)},
\end{equation}
which leads to 
\begin{equation}
    \int_{a}^{b} \tilde{Y}_t \, d\tilde{X}_t = \int_{a}^{b} Y_{\psi(t)} \dot{\psi}(t) \,d X_{\psi(t)}  = \int_{a}^{b} Y_u \, dX_u,
\end{equation}

The equivalence of the two integrals is obviously obtained through the change of variable 
$ u := \psi(t) $. This confirms the invariance of path integrals with respect to time reparametrization of the paths involved (\cite{chevyrev2016primer,lyons2014rough}). Let us consider a multi-dimensional path $X:[a,b]\to \Rb^{d} $ and a reparametrization $ \psi: [a, b] \to [a, b] $. As before, denote by $ \tilde{X}: [a, b] \to \Rb^d$ the reparametrized path $ (\tilde{X}_t) = (X_{\psi(t)})$. Since every term of the signature  $S(X)_{a,b}^{i_1,\ldots,i_k}$  is defined as an iterated path integral of $(X_t)$, it follows from above that

\begin{equation}
    S(\tilde{X})_{a,b}^{i_1,\ldots,i_k} = S(X)_{a,b}^{i_1,\ldots,i_k}, \quad \forall k \geq 0, \, i_1, \ldots, i_k \in \{1, \ldots, d\}.
\end{equation}

That is to say, the signature $ S(X)_{a,b}$ remains invariant under time reparametrizations of $(X_t)$. 

\subsection{Signature of a stream}

Even if financial time series are always recorded at discrete interval in practice (closing times, opening times, etc.), it may be considered that such time series evolve continuously in time. Due to its convenience on the theoretical and practical sides, continuous time models are very often used to represent the dynamics of financial assets. Indeed, dealing with continuous intervals implies the use of data streams, sequences of data points $X:=(X_1,...X_N)$ associated to a timestamp. In the previous sections, we have seen that the signature $S(X)$ of $(X_t)$ is invariant to time reparameterization. However, in some fields, such as economics and finance, there is a need for time-reparameterization of the path $(X_t)$. 

\medskip

Let $(\widehat{X_{t_i}})_{i=0}^N$ be a data stream. Each point is associated with a specific timestamp. $X_{t_i}$  is the value of the path at time $t_i$. The objective is to transform this discrete data stream into a continuous path. This transformation can be achieved through a transformation that "connects" all the data points over time.  The time-joined transformation \cite{levin2013learning}, or the lead-lag transformation allows to perform this "connection". \cite{gyurko2013extracting}  demonstrated that the terms of the path signature (iterated integrals) quantify different path-dependent attributes. This method does not  account for the quadratic variation of the process which is a very important metric in finance. So they achieved the integration of the quadratic variation using the lead-lag transformation of data streams. 
Given a stream  $(\widehat{X}_{t_i})_{i=0}^N \in \mathbb{R}^d$, we define its lead-transformed stream $(\widehat{X}^{\text{lead}}_j)_{j=0}^{2N}$ by
\[
\widehat{X}^{\text{lead}}_j=\left\{
\begin{array}{ll}
\widehat{X}_{t_i} & \text{ if } j = 2i, \\
\widehat{X}_{t_{i}} & \text{ if } j = 2i-1.
\end{array}
\right.
\]
Moreover, we define its lag-transformed stream $(\widehat{X}^{\text{lag}}_j)_{j=0}^{2N}$ by
\[
\widehat{X}^{\text{lag}}_j=\left\{
\begin{array}{ll}
\widehat{X}_{t_i} & \text{ if } j = 2i, \\
\widehat{X}_{t_{i}} & \text{ if } j = 2i+1. 
\end{array}
\right.
\]
Finally, the lead-lag-transformed stream takes values in
$\mathbb{R}^{2d}$ and is defined by the paired stream
\[
(\widehat{X}^{\text{lead-lag}}_j)_{j=0}^{2N} = (\widehat{X}^{\text{lead}}_j,\widehat{X}^{\text{lag}}_j)_{j=0}^{2N}.
\]

Note that the axis path $X^{\text{lead}}$ corresponding to the
lead-transform stream is a time-reparametrization of the axis path
$X$, hence 
\[
S_{t_0,t_N}(\widehat{X}^{\text{lead}}) = S_{0,2N}(\widehat{X}^{\text{lead}}),
\]
and similarly 
\[
S_{t_0,t_N}(\widehat{X}^{\text{lag}}) = S_{0,2N}(\widehat{X}^{\text{lag}}).
\]

Moreover, the (signed) area spanned between the $i$th element of the lead-transform and the $j$th element of the lag-transform is equivalent to the quadratic cross-variation of the paths $\widehat{X}^i$ and $\widehat{X}^j$.

\section{Data and Methodology}

In the first step, we select the largest digital assets in terms of market capitalization. Thus, we are confident that the liquidity of each digital asset is sufficient to replicate the strategy in the real world. Since some recently created digital assets listed on the market may face liquidity issues due to their youth, these digital assets should be excluded from this initial investment universe.  Moreover, the most liquid digital assets suffer from less frequent price big jumps than the others. They will not be subject to significant slippage effects when trades are executed, which helps preserve an acceptable tracking error for asset managers.

\medskip
After identifying this list of digital assets, the objective is to detect connections and/or similarities between the dynamics followed by different digital assets, to create homogenous clusters based on a certain ``intrinsic'' characteristics of their price dynamics. To perform this task, we rely on the k-means algorithm, that is applied to the path signatures of each of the selected digital assets.

\begin{figure}[H]
    \centering
    \includegraphics[width=0.6\columnwidth]{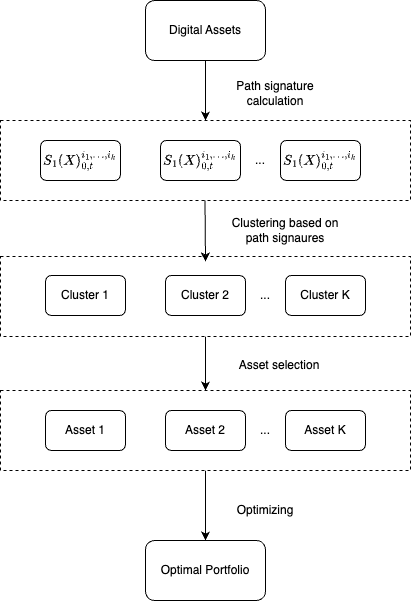}
    \caption{Portfolio construction methodology}
    \label{fig:building_optimal_portfolio_using_sig}
\end{figure}

Figure \ref{fig:building_optimal_portfolio_using_sig}  describe the four steps of our methodology. Our steps are similar to the one of \cite{manh2023portfolio}, however our initial dataset "Digital Assets"  will be used as a basis for  path signatures calculation, which will be the features of the clustering algorithm. Our portfolio construction flow may be described in four distinct steps. The first step is to calculate the path signatures for each assets within the initial investment universe. The second step consists in grouping digital assets according their path signatures. During the third step, digital assets will be selected according to their distance to the centroid of the cluster. For the final step the optimization algorithm will be applied to the digital assets selected from the third step.

\medskip

In our experiment, we used a dataset downloaded from Binance. This dataset includes the top 30 digital assets in terms of market capitalisation, according to \hyperlink{https://coinmarketcap.com/}{Coinmarketcap}.
We downloaded the daily closing price data for all the trading pairs "crypto against USD", and use USDT as a proxy of the US dollar. We only keep the closing prices for each pair, observed at midnight every day. From this data set, we create an initial investment universe at each rebalancing date. In our case, rebalancing events will be set up at midnight every monday. The rebalancing date is the moment at which the filtered universe is built on top of the initial investment universe. As the number of digital assets tends to grow, some did not yet exist at the start of our backtest. These assets will be added to the investment universe as time goes thanks to our rebalancing mecanism. 

\medskip

For the backtest methodology we will use to types of lookback windows. One is called the Fixed Origin of Time (FOT) method and the other is the traditional rolling window (RW) strategy. The FOT window is growing with time while the rolling window keeps the same size during the whole backtest. The relationships between digital assets evolve over time. We will analyze whether a "static" time frame called the FOT is more reliable than a rolling window, or vice versa, for developing a new portfolio methodology.

\subsection*{Fixed Origin of Time (FOT)}
The Fixed Origin of Time (FOT) window methodology is characterized by its progressive expansion over time, enabling the integration of an increasing amount of historical data. The cluster algorithm will incorporate each reshuffle date. We operated on a sequential basis. We downloaded the time series for each asset, focusing solely on their closing prices. Then, we generated a set of dates. We identified the crypto assets that are available at each date, allowing us to construct our investment universe. Then, we gradually expanded our universe of digital assets  with the introduction of new ones over time. Let us define $N_t$ as the number of crypto-asset at time $t$. The vector of path signature $Q_t$, of the $n$ asset at time $t$, $n \in [1,...,N_t] $, is given by

\begin{equation}
    Q_t := [S_1(X)_{t_0,t}^{i_1,\ldots,i_k},S_n(X)_{t_0,t}^{i_1,\ldots,i_k},...,S_{N_t}(X)_{t_0,t}^{i_1,\ldots,i_k}],
\label{def_QI}
\end{equation}
Where $S_n(X)_{0,t}^{i_1,\ldots,i_k}$ is the the $k$-th level of path signatures of the $n$-th digital asset over the interval $[0,t]$. This vector will be the input of the clustering algorithm.

\medskip

For clustering purpose, we choose the k-means algorithm which aims to divide a set of $m$ points $( x_1, x_2, \ldots, x_m )$ into $k$ groups, each described by the mean of the points in the group. The objective function to be minimized is defined as the sum of the squared distances between each data point and its assigned cluster mean. Given $\{x_i\}_{i=1}^m \in \Rb^d$, the objective is to identify the centroids $\{c_j\}_{j=1}^k$ of the clusters and to minimize the distance between each data point and its centroid. The $k$ means loss function is
\begin{equation}
    \Lc(c_1,\Gamma_1,\ldots,c_k,\Gamma_k):= \sum_{j=1}^{k} \sum_{i\in \Gamma_j} d(x_i,c_j),
    \label{eq:loss_k_means}
\end{equation}
where $\Gamma_j$ denotes the $j$-th cluster, whose centroid is $c_j$. We denoted $d(x_i,c_j)$ a distance between the data point $x_i$ and the centroid of the $j$-th cluster. Usually an Euclidian distance is used in this problem, i.e. $d(x_i,c_j)=||x_i-c_j ||^2$. 

\medskip

Let us denote $\{c_{j}^{s}\}_{j=1}^k$ our centroids at time $t$.
Once $k$ clusters have been built, we assign each data point to its closest centroid. 
Here, the "closest" is determined by the Euclidean distance between a point and a centroid. 
This \emph{assigment step} can be represented as
\begin{equation}
    \Gamma_{j}^s:= \{ i: ||x_i-c_{j}^s|| \leq ||x_i-c_{k}^s||,\; \forall k \neq j \}.
\end{equation}
For a given cluster $ \Gamma_{j}^s$, the minimizer of (\ref{eq:loss_k_means}), the centroid of $\Gamma_{j}$ is \emph{updated} as follows:

\begin{equation}
    c_{j}^{s+1} = \argmin_{c_j} \sum_{i \in \Gamma_{j}^{s}} ||x_i - c_j||^2=\frac{1}{| \Gamma_{j}^s |}\sum_{i\in \Gamma_{j}^s} x_i.
\end{equation}
The operation is repeated until the convergence of the algorithm when centroids dot not change significantly. To illustrate you will find below a picture of the cluster for a given $t$.

\medskip

\begin{figure}[H]

\begin{subfigure}{0.5\textwidth}
\includegraphics[width=1\linewidth, height=6cm]{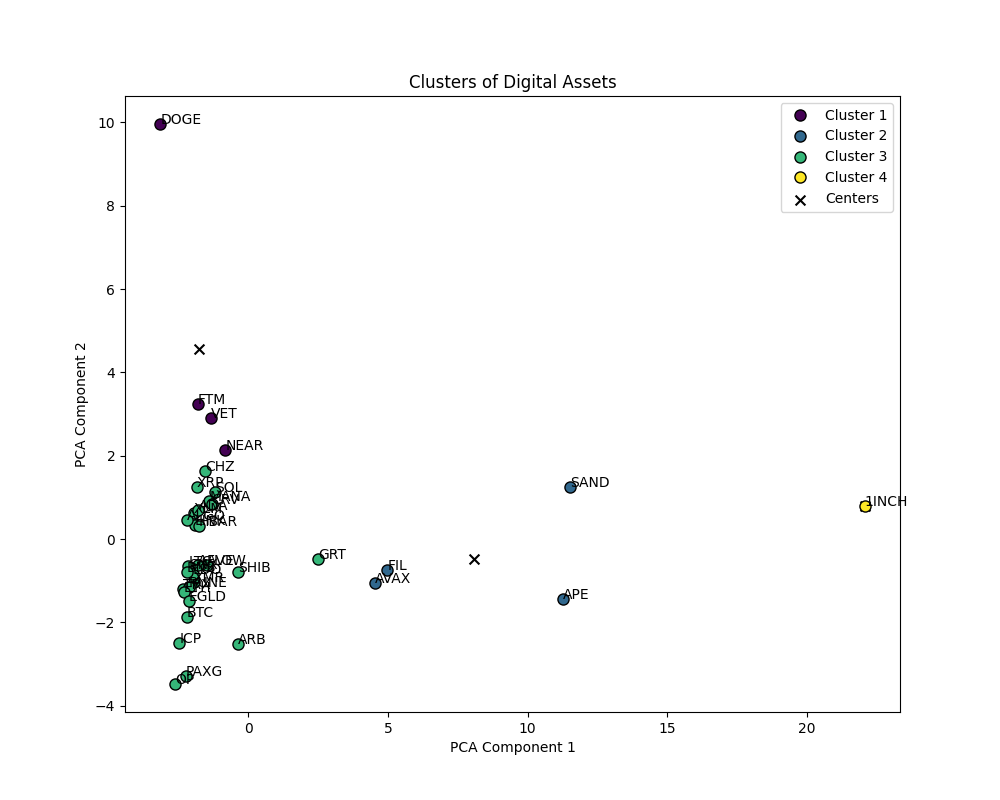} 
\caption{FOT Clusters}
\label{fig:subim1_fot}
\end{subfigure}
\begin{subfigure}{0.5\textwidth}
\includegraphics[width=1\linewidth,height=6cm]{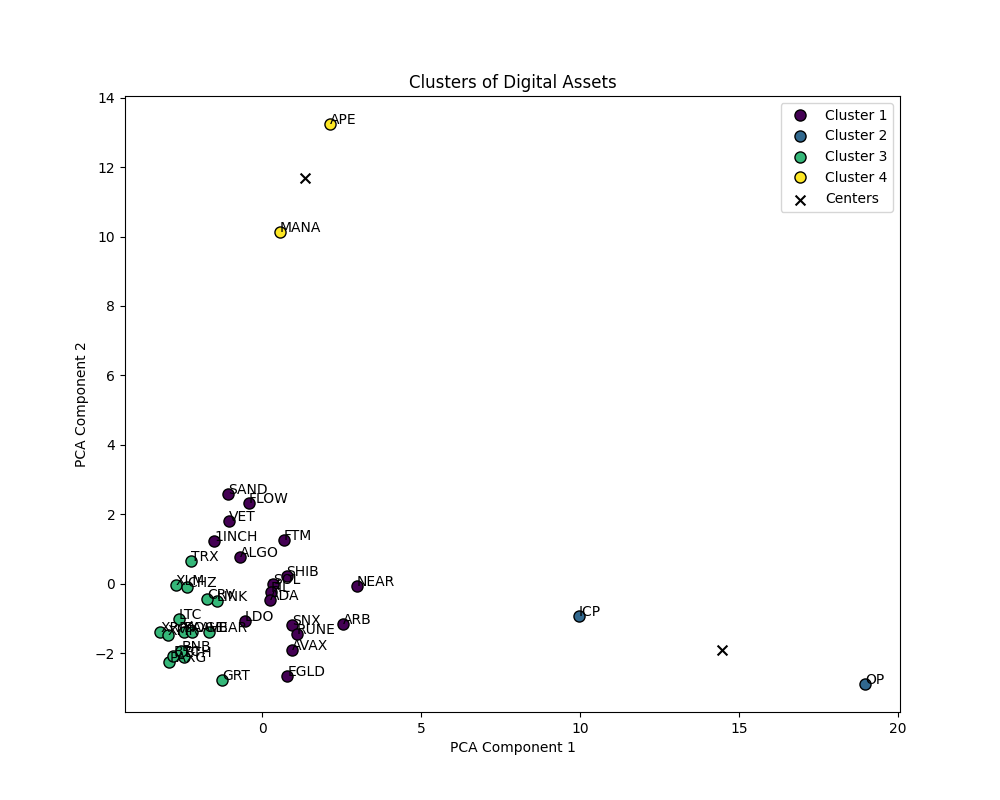}
\caption{RW Clusters}
\label{fig:subim1_rw}
\end{subfigure}
 
\caption{Comparison of FOT and RW clusters as of 2023-12-25.}
\label{fig:image2}
\end{figure}

\begin{figure}[H]
    \centering
\includegraphics[width=1.\columnwidth]{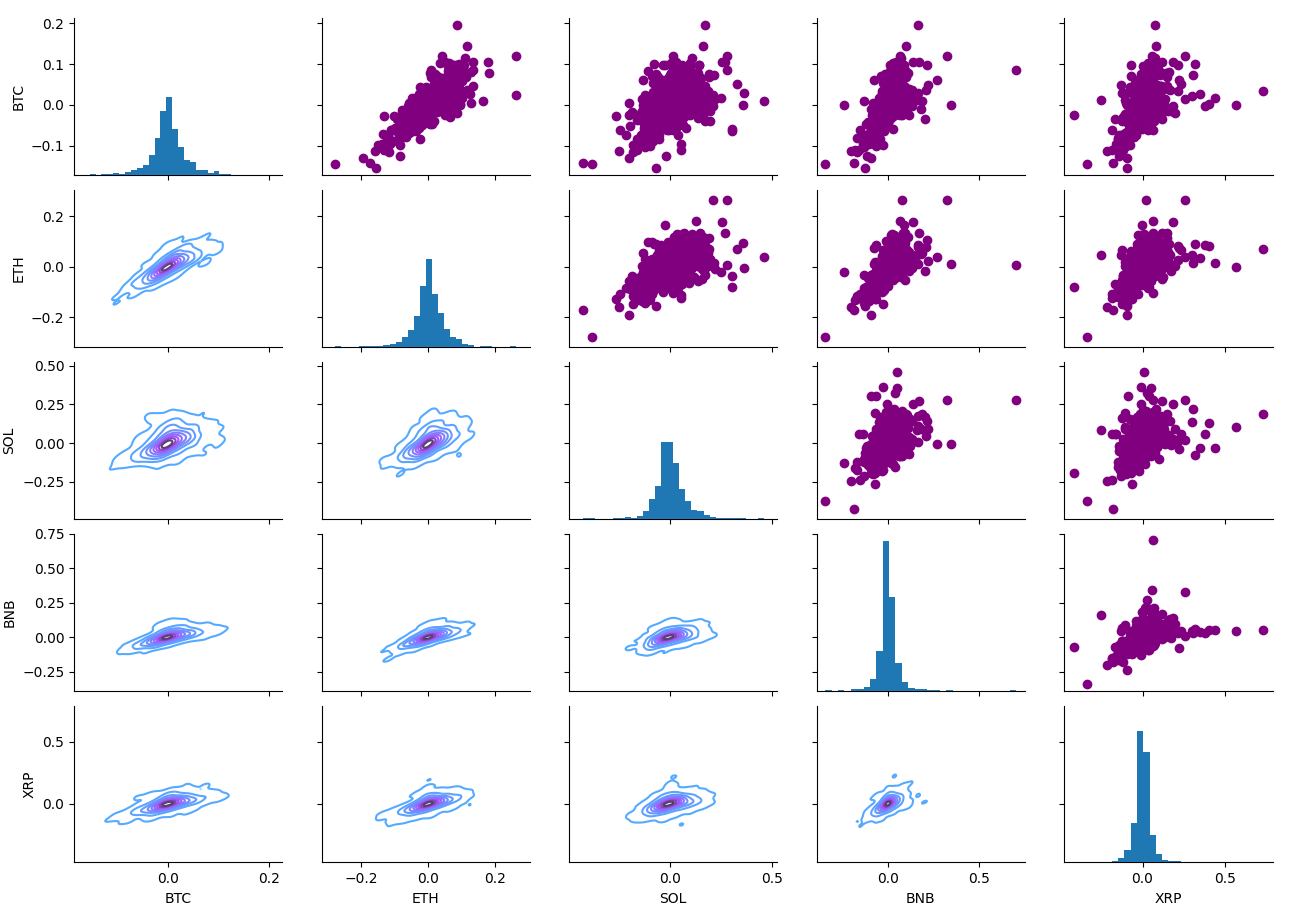}
\caption{Scatterplots, histograms and joint distributions for the components of Cluster 3 (FOT).}
\label{fig:crypto_statistics}
\end{figure}

In Figure \ref{fig:subim1_fot}, we plotted the clusters obtained with the FOT methodology at some arbitrarily chosen date. It can be observed that the majority of digital assets including BTC, ETH, and LTC are very close to each other in  cluster 2 (in blue), which indicates similarities in terms of path signatures of these digital assets. Digital assets such as SAND, APE and DOGE are located far from the previous points in the chart, which could indicate a more marked difference from the others in terms of characteristics and statistical behavior.  1INCH also differed from the other assets even more strikingly. It seems to have a different behavior. This asset could probably bring an interesting amount of diversification inside a digital asset portfolio. To build clusters using the (FOT) method, we keep a complete history of assets, which is fed continuously as time goes. This means that some assets may not exist at the start, but are added throughout the backtest. This methodology ensures greater stability over time within the clusters. Some unselected digital assets could have been subject to a breach of the exchange platform, a hack or other threats leading to their disappearance. Cluster 2 gathers the largest digital assets in term of market cap. We supposed the latter digital assets move in the save direction ``in general'' (without any special market condition, such as some news related to a specific digital asset typically). Figure \ref{fig:crypto_statistics} gives us an overview of the statistical behavior of the digital assets within the largest cluster (Cluster 1 \ref{fig:subim1_fot}). The scatterplots of log returns show that these digital assets are positively dependent. The distributions illustrated by the histograms vary according to each digital assets: we observe histograms with high and sharp peaks which indicates low volatility, and others with wider bases, a sign of commonly observed high volatility. As for the contours, they indicate that for some pairs, there are areas where returns are particularly concentrated, which could signify similar market behaviours. 

\subsection*{Rolling Window (RW)}

The use of the sliding window is particularly relevant in our context of cluster construction at each rebalancing date. At each rebalancement date, the signatures of the existing assets are calculated on the (past) rolling window period of time. The assets that appeared since the last rebalancement date are then progressively included in the universe of assets. The advantage of this method is its flexibility and its ability to rapidly integrate market changes. By using a sliding window, we assume that recent past price behaviors offer the best indications of future trends. Our objective is to determine whether the rolling window, with its responsiveness to market conditions, provides a more solid insight for portfolio construction than the Fixed Origin of Time window which incorporates a longer historical perspective. This comparison will be made in a latter section, which aims to identify the strategy offering the best balance between taking long-term trends into account and reacting to recent market developments to optimize portfolio performance. 

\medskip

For the Rolling Window (RW) methodology, we maintain a consistent window size of thirty days throughout the backtesting period. Unlike the Fixed Origin of Time (FOT) methodology, which includes more and more historical data (through prices) over time, the rolling window method moves forward in time, keeping the quantity of analyzed data constant but updating it to reflect the most recent price patterns. We define $N_t$ as the number of crypto assets available in our investment universe at time $t$ within the rolling window, and $Q_t, n \in [1,...,N_t]$ as the vector representing the path signatures of the $n$-th asset within this window. $Q_t$ is given by:  

\begin{equation}
    Q_t := [S_1(X)_{t-q,t}^{i_1,\ldots,i_k},S_n(X)_{t-q,t}^{i_1,\ldots,i_k},...,S_{N_t}(X)_{t-q,t}^{i_1,\ldots,i_k}],
\label{def_QI_rolling}
\end{equation}
where $S_n(X)_{t-q,t}^{i_1,\ldots,i_k}$ is the path signature of the $n-th$ asset over the time window $[t-q,t]$. 

This methodology allows for a dynamic adjustment to our portfolio's asset composition. It ensures that our investment decisions are based on the most recent and relevant data, thus striving to enhance our portfolio's performance by adapting to the most recent trends in the market.

\medskip

Figure \ref{fig:subim1_rw} show the different clusters using a rolling window of thirty days. 
Comparing the two methodologies, the RW clusters seem to be more separated, especially for Cluster 2 which is more dispersed. The RW clusters appear to have a tighter central cluster (Cluster 1) and less dispersion in Cluster 2. In both methods, Cluster 3 contains a few assets only. Cluster 4 has got even fewer, and they both appear to have outliers that are far from the cluster centers. The scale on the PCA Component 1 axis is larger for the FOT Clusters than the RW Clusters, indicating a broader distribution of data in the FOT case. Nonetheless, these results are related to a particular day and it is difficult to assess generalities.

\section{Application}

For our comparative study of digital asset portfolio strategies, we built three different portfolios: equally weighted, mean variance (\cite{markowitz2000mean}) and maximum diversification (\cite{choueifaty2008toward}) portfolios. 
For our application, our objective is to determine if portfolios refined through clustering techniques can outperform more traditional investment strategies in performance. 
We started with a portfolio investment universe without imposing any selection criteria. Then, we calculate the portfolio value for each date. We refine our investment universe by retaining only digital assets nearest to the centroids of each defined class. This approach aims to lower rebalancing costs by selecting only one asset. This approach seeks to reduce portfolio rebalancing costs, which should be lower for portfolios built using clustering than for those on which no filter has been applied. It is expected that portfolios with fewer assets will exhibit relatively higher volatility compared to more diversified ones.

\medskip

In this section, we compare the performances of the filtered and not-filtered strategies in the case of equally weighted portfolio, the mean-variance portfolio and the maximum diversification portfolio. Our comparisons will be made firstly on the performance of the strategies and secondly on some performance metrics. For each portfolio, we assume we are able to execute rebalancing at closing times (00:00 UTC). The rebalancing frequency is the same for every strategy; the single difference relais on the allocation model that is used and on the upstream clustering that creates a more parsimonious investment universe than was initially planned. 

\subsection{Equally Weighted Portfolio (EW)}

Below, Table \ref{table:ew_perf} presents a comparison between the annualized return (resp. annualized volatility) of two portfolios: the equally weighted portfolio $(EW)$ against the equally weighted  portfolios with clustering filters $(EW_{sc}^{FOT})$ or $(EW_{sc}^{RW})$. 
$(EW_{sc}^{FOT})$  has a higher annualized return of 0.9592, indicating  better performance and also a lower annualized volatility of 0.6319, which imply it has less risk in terms of the variability of its returns. $(EW_{sc}^{RW})$ has the highest annualized return of 1.2523, which indicates a more successful strategy in terms of returns, certainly due to the rolling window which during the reshuffle has allowed the selection of some assets that perform better (under a ``momentum'' perspective).

    \begin{table}[H]
        \centering
        \caption{Performance EW Portfolios (FOT)}
        \begin{tabular}{lcc}
        \toprule
            ~ & Annualized Return & Annualized Volatility  \\ \midrule
            PORTFOLIO\_EW & 0.5984 & 0.8219  \\ 
            PORTFOLIO\_SIG\_CLUSTER\_EW\_FOT & 0.9592& 0.6319 \\ 
             PORTFOLIO\_SIG\_CLUSTER\_EW\_RW & 1.2523& 0.7432 \\ 
             \bottomrule
        \end{tabular}
        \label{table:ew_perf}
    \end{table}

    \begin{table}[H]
         
        \centering
        \caption{Risk metrics EW Portfolios (FOT)}
        \begin{tabular}{lccc}
        \toprule
            ~ & Sharpe & Calmar & MDD \\ \midrule
            PORTFOLIO\_EW\ & 0.7281 & 0.7291 & 0.8203  \\ 
            PORTFOLIO\_SIG\_CLUSTER\_EW\_FOT & 1.5178 & 1.5879 & 0.6036 \\ 
             PORTFOLIO\_SIG\_CLUSTER\_EW\_RW & 1.6849 & 2.0306 & 0.6163 \\
             \bottomrule
        \end{tabular}
        \label{table:ew_risk}
    \end{table}
Table \ref{table:ew_risk}, shows some performance metrics of the latter portfolios. For $(EW_{sc}^{RW})$, we observe a relatively high Sharpe and Calmar ratios (1.6849 and 2.0306, respectively). This indicates a better risk-adjusted return compared to $(EW_{sc}^{FOT})$ and $(EW)$. Signature-based clustered stratgies exhibit lower Maximum DrawDown (MDD) than $(EW)$, suggesting that the former strategies have 
experienced smaller peak-to-trough declines during our studied period, which is desirable in terms of risk management. 
Globally, $(EW_{sc}^{RW})$ appears to outperform $(EW_{sc}^{FOT})$ and  $(EW)$  in terms of both return and risk metrics, suggesting that it might be the most efficient option for investors who want to diversify their portfolio, despite the riskness of this asset class. There is also a lower maximum capital loss on the portfolio based on the filtered universe. This is an advantage for decision-making, even if past performance is no guarantee of future performance.

\subsection*{Fixed Origin of Time (FOT)}
 \textit{In this subsection you will find figures of the evolution of the portfolio allocation (weights) as well as the portfolio values backtested with the methodology that uses the Fixed Origin of Time window (FOT)}
\subsubsection{Without Clustering}
    \begin{figure}[H]
        \centering
    \includegraphics[width=1.\columnwidth]{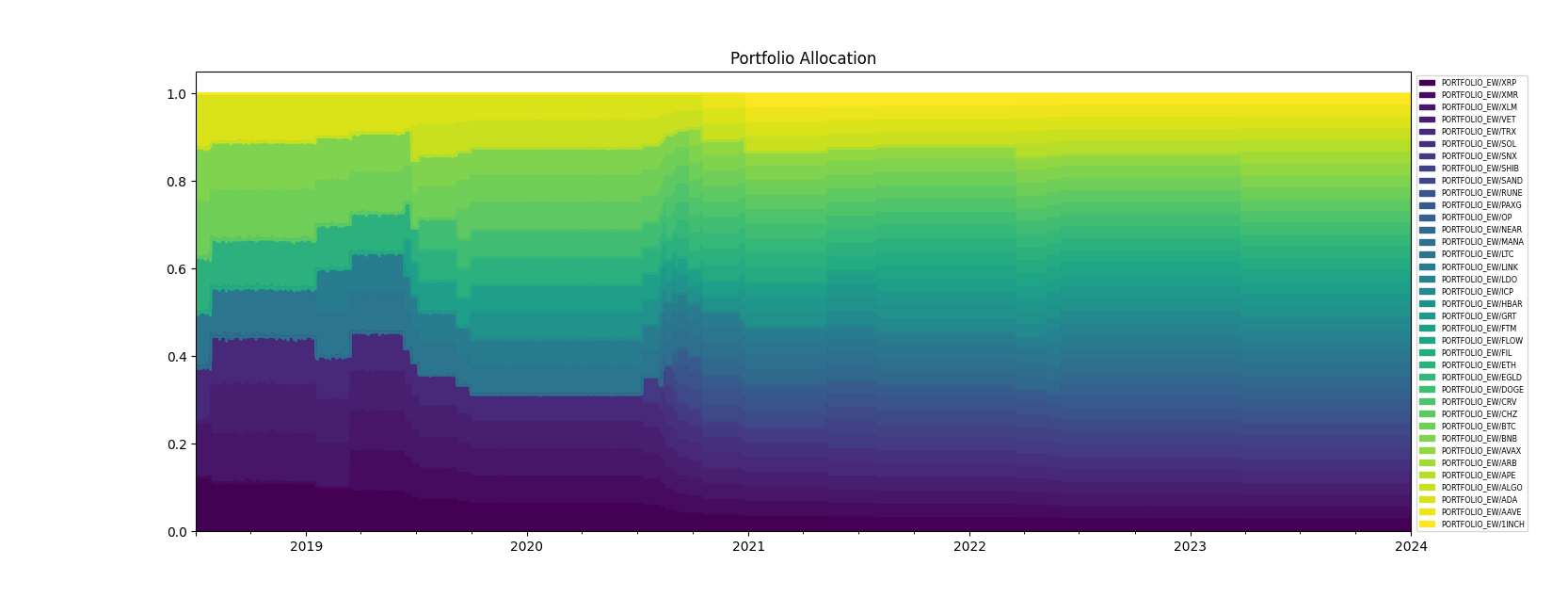}
    \caption{EW Portfolio allocation (FOT)}
    \label{fig:EW_Weights}
    \end{figure}

\subsubsection{With Clustering}

    \begin{figure}[H]
        \centering
    \includegraphics[width=1.\columnwidth]{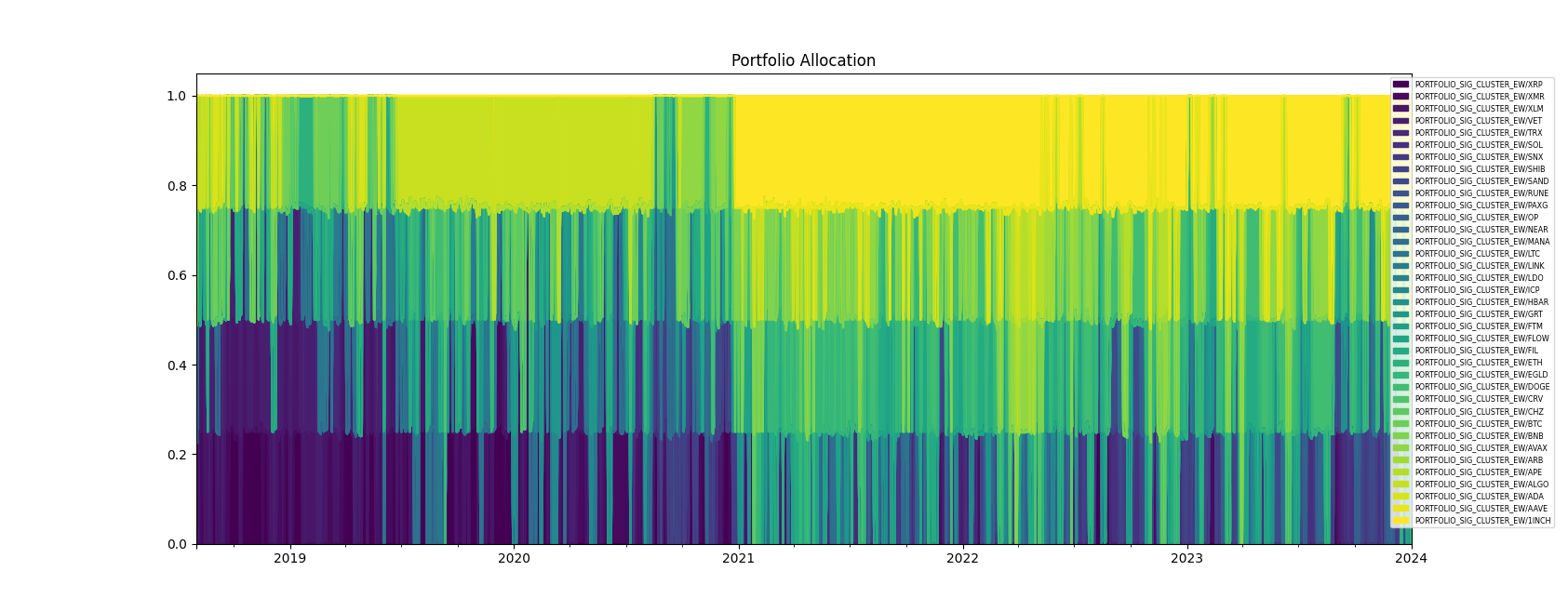}
    \caption{Signature Clustered EW Portfolio allocation (FOT)}
    \label{fig:EW_Cluster_SIG_Weights}
    \end{figure}

\begin{figure}[H]
    \centering
    \includegraphics[width=1\columnwidth]{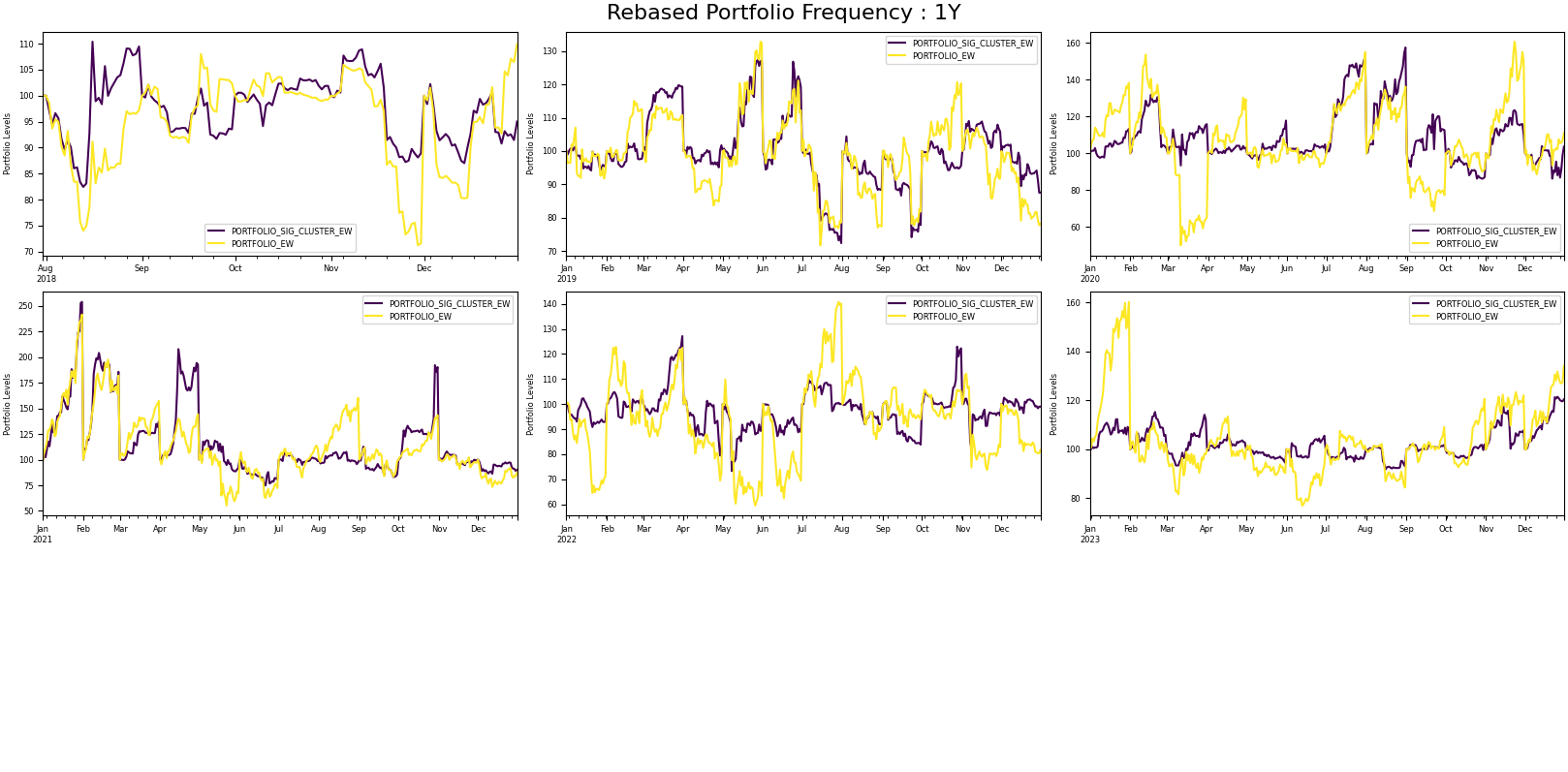}
    \caption{EW Portfolio Rebased Annually (FOT)}
    \label{fig:EW_Rebased_Portfolio}
    \end{figure}

Figures \ref{fig:EW_Weights} - \ref{fig:EW_Cluster_SIG_Weights} illustrates the changes in portfolio allocation throughout the backtest period. Specifically, Figure \ref{fig:EW_Cluster_SIG_Weights} perceptibly highlights the introduction of new assets into the portfolio composition over time. Figure \ref{fig:EW_Rebased_Portfolio} shows the value of the portfolios over time. We can clearly see that, with annual rebasing, the $(EW_{sc}^{FOT})$ portfolio is the least volatile.

\subsection*{Rolling Window (RW)}
    \textit{In this subsection you will find additional figures of the evolution of the portfolio allocation as well as the portfolio values backtested with the methodology that uses the rolling window (RW)}

  \subsubsection{Without Clustering}

        \begin{figure}[H]
            \centering
        \includegraphics[width=1.\columnwidth]{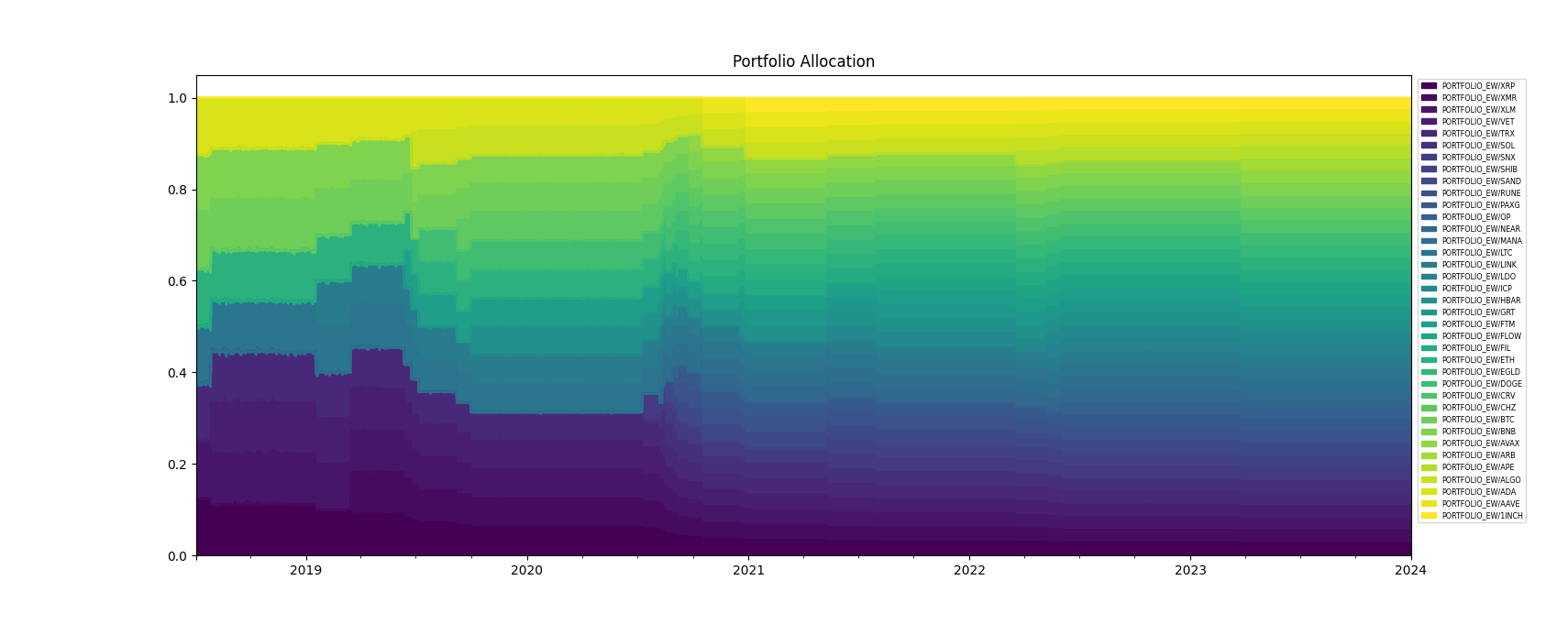}
        \caption{EW Portfolio allocation (RW)}
        \label{fig:EW_Weights_RW}
        \end{figure}

    \subsubsection{With Clustering}
        
        \begin{figure}[H]
            \centering
        \includegraphics[width=1.\columnwidth]{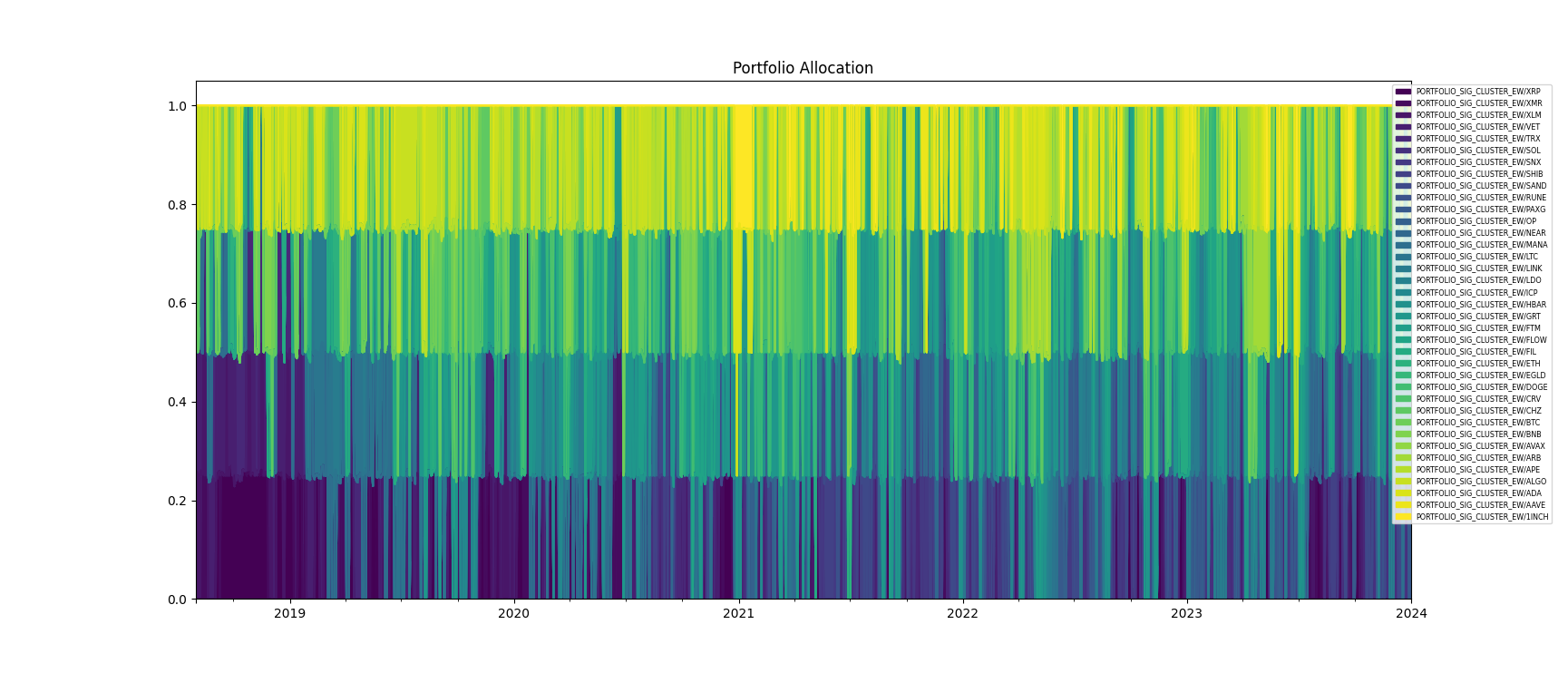}
        \caption{Signature Clustered EW Portfolio allocation (RW)}
        \label{fig:EW_Cluster_SIG_Weights_RW}
        \end{figure}

        \begin{figure}[H]
        \centering
        \includegraphics[width=1\columnwidth]{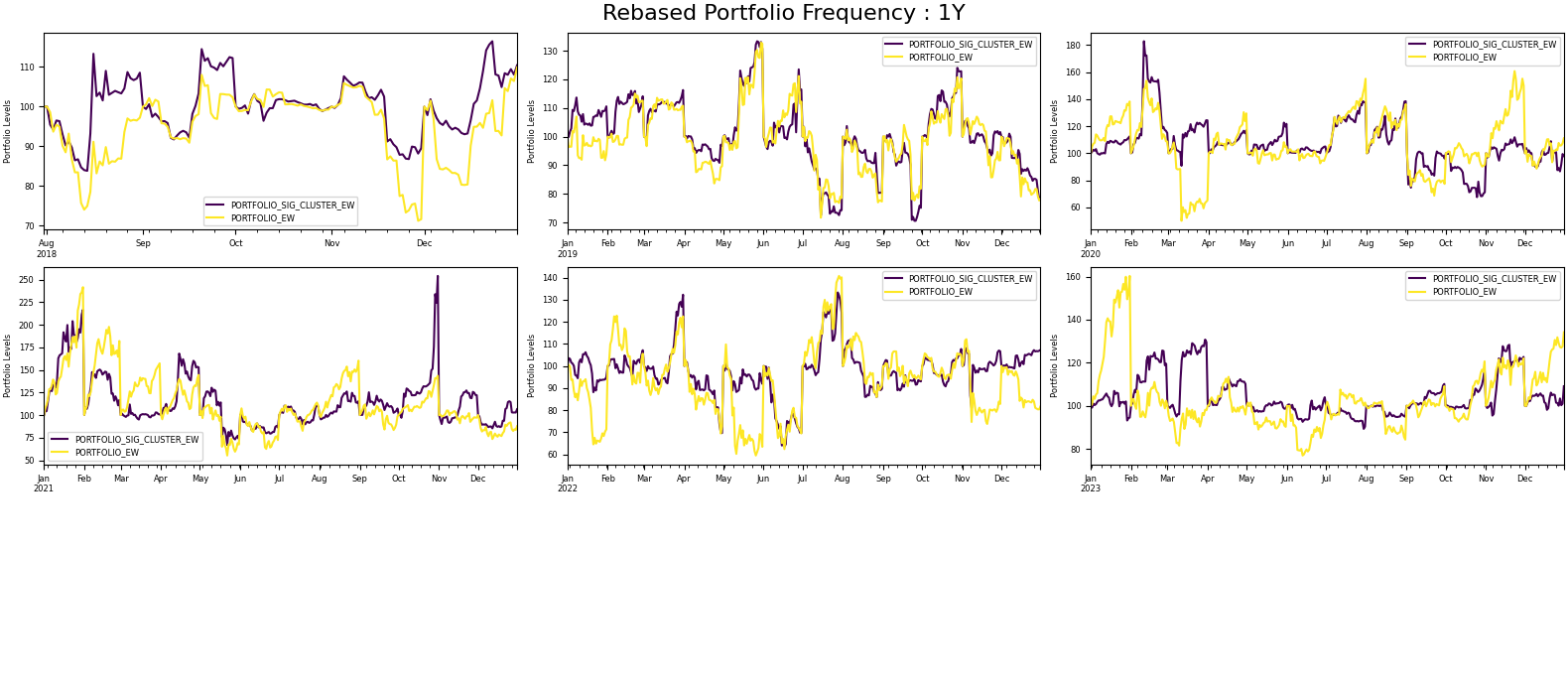}
        \caption{EW Portfolio Rebased Annually (RW)}
        \label{fig:EW_Rebased_Portfolio_RW}
        \end{figure}

The next backtesting exercice concerns the MVP strategy, a model where volatility is at the heart of the optimization program. 
We hope that a reduction of the portfolio size can have a strong impact on the overall portfolio volatility.

\subsection{Mininimum Variance Portfolio (MVP)}

Table \ref{table:mvp_perf} compares the annualized return and annualized volatility of two portfolios: the mean variance portfolio $(MVP)$ and mean variance portfolio after clustering filter $(MVP_{sc}^{FOT})$ or $(MVP_{sc}^{RW})$. The strategy $(MVP_{sc}^{FOT})$ shows an annualized return of 0.2199, indicating better performance than the competitiors, but at the price of a higher level of annualized volatility (0.6775 against 0.4262 for $(MVP)$). When we look at $(MVP_{sc}^{RW})$, it seems that adapting to market changes and trends using the rolling window is not a good strategy in this case.

        \begin{table}[H]
        \centering
        \caption{Performance MV Portfolios}
        \begin{tabular}{lcc}
        \toprule
            ~ & Annualized Return & Annualized Volatility  \\ \hline
            PORTFOLIO\_MVP & 0.1140 & 0.4262  \\ 
            PORTFOLIO\_SIG\_CLUSTER\_MVP\_FOT &0.2199 & 0.6775 \\ 
            PORTFOLIO\_SIG\_CLUSTER\_MVP\_RW &0.1488 & 0.6675 \\ 
        \bottomrule
        \end{tabular}
                \label{table:mvp_perf}
    \end{table}

    \begin{table}[H]
        \centering
        \caption{Risk metrics MV Portfolios}
        \begin{tabular}{lccc}
        \toprule
            ~ & Sharpe & Calmar & MDD \\ \hline
            PORTFOLIO\_MVP & 0.2674 & 0.2510 &0,4539  \\ 
            PORTFOLIO\_SIG\_CLUSTER\_MVP\_FOT & 0.3245 & 0.3171 & 0.6928 \\
            PORTFOLIO\_SIG\_CLUSTER\_MVP\_RW & 0.2229 & 0.1754 & 0.8476 \\ 
        \bottomrule
        \end{tabular}
        \label{table:mvp_risk}
    \end{table}

  Table \ref{table:mvp_risk} shows the risk metrics for both portfolios, $(MVP_{sc}^{FOT})$ shows slightly higher Sharpe and Calmar ratios of 0.3245 and 0.3171, respectively. In other words, a better risk-adjusted return compared to $(MVP)$. However, the maximum drawdown of the portfolio with clustering filter is higher than the one with no filter. It seems that the representative asset of the class picked during the portfolio universe construction is catching the most deterministic trend.  Figures \ref{table:mvp_perf}-\ref{table:mvp_risk} suggest that $(MVP_{sc}^{FOT})$ has a better risk-adjusted return than $(MVP)$ portfolio. Considering $(MVP_{sc}^{RW})$, we might infer that this methodology is not optimal for minimizing the portfolio variance. Utilizing a rolling window approach yields in a lower sharpe ratio of 0.2239, indicating that $(MVP_{sc}^{RW})$ performs worse compared to $(MVP)$. Therefore, applying a filter on the investment universe in this case may not be effective. For risk management purpose, it is important to note that $(MVP_{sc}^{FOT})$ remains a more risky portfolio.

\subsection*{Fixed Origin of Time (FOT)}
 \textit{In this subsection you will find figures of the evolution of the portfolio allocation as well as portfolio values backtested with the methodology using the Fixed Origin of Time window (FOT)}
    \subsubsection{Without Clustering}

        \begin{figure}[H]
            \centering
        \includegraphics[width=1.\columnwidth]{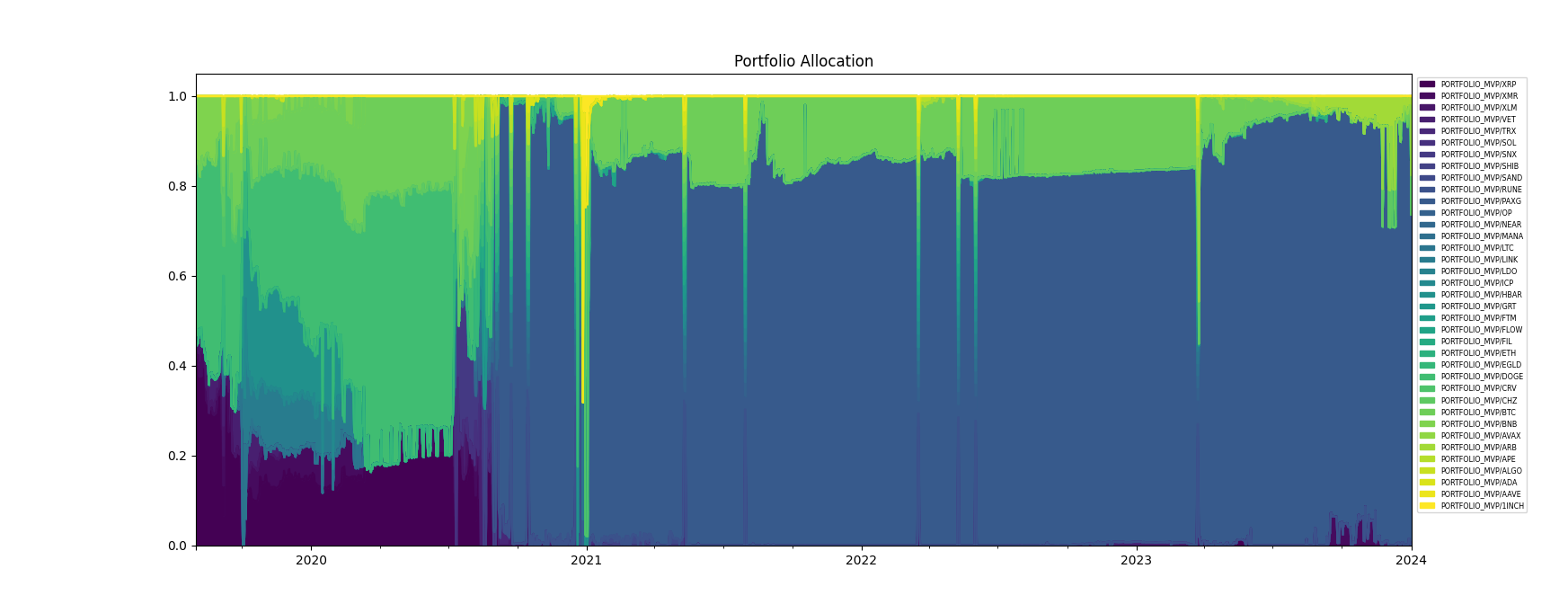}
        \caption{MVP Portfolio allocation (FOT)}
        \label{fig:MVP_Weights}
        \end{figure}

    \subsubsection{With Clustering}
        
        \begin{figure}[H]
            \centering
        \includegraphics[width=1.\columnwidth]{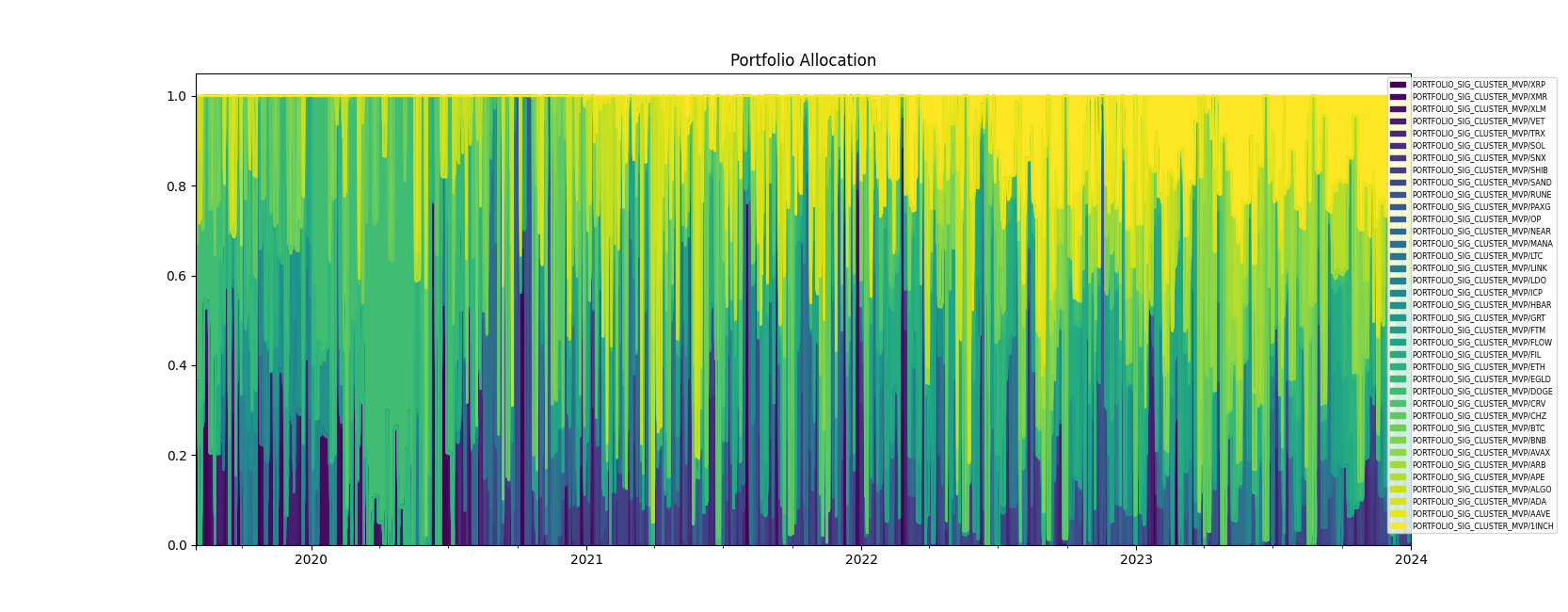}
        \caption{Signature Clustered MVP Portfolio allocation (FOT)}
        \label{fig:MVP_Cluster_SIG_Weights}
        \end{figure}

        \begin{figure}[H]
        \centering
        \includegraphics[width=1\columnwidth]{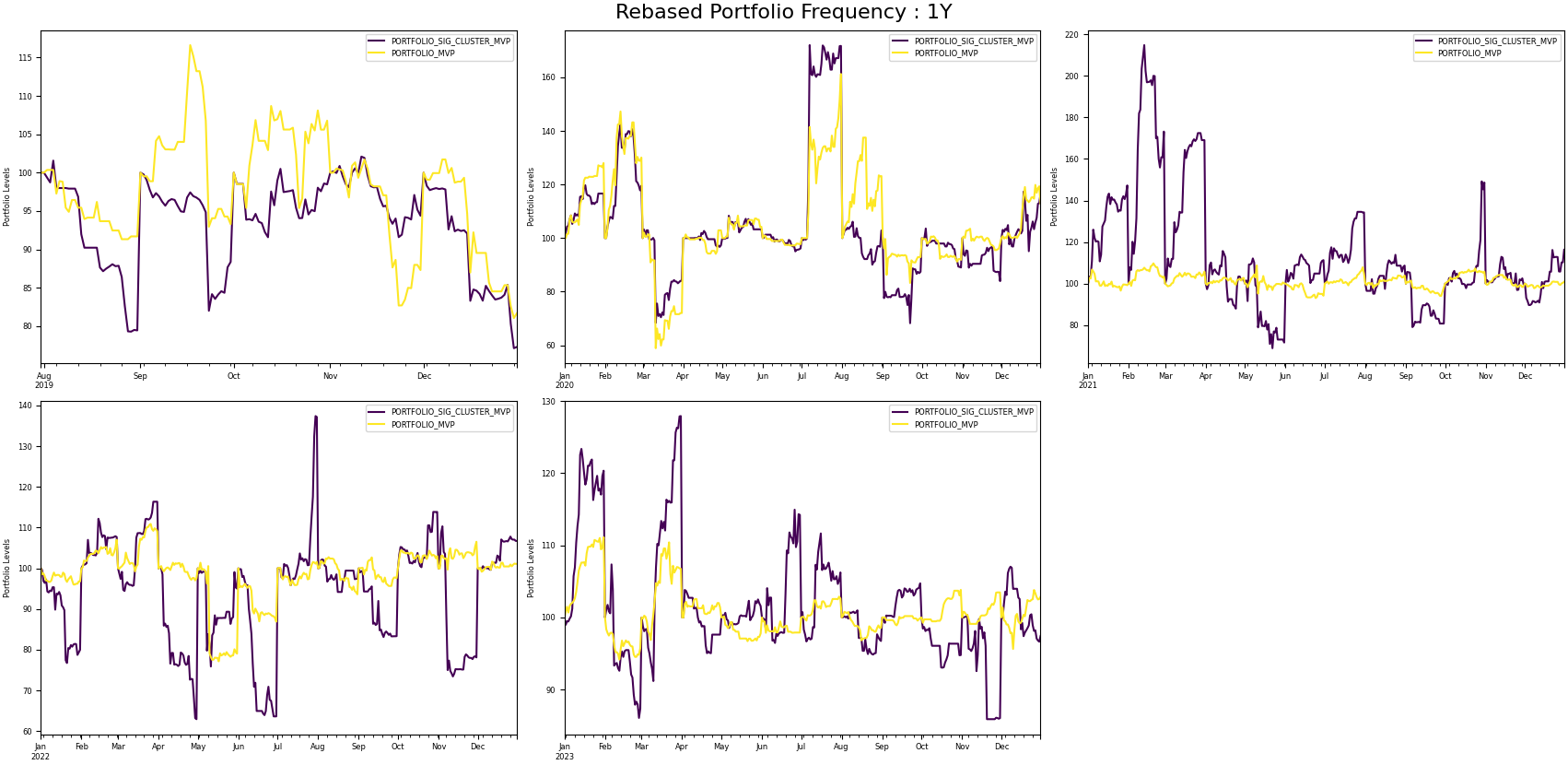}
        \caption{MVP Portfolio Rebased Annually (FOT)}
        \label{fig:MVP_Rebased_Portfolio}
        \end{figure}

\subsection*{Rolling Window (RW)}
    \textit{In this subsection you will find additional figures of the evolution of the portfolio allocation as well as the value of portfolios backtested with the methodology using the rolling window (RW)}
  \subsubsection{Without Clustering}

        \begin{figure}[H]
            \centering
        \includegraphics[width=1.\columnwidth]{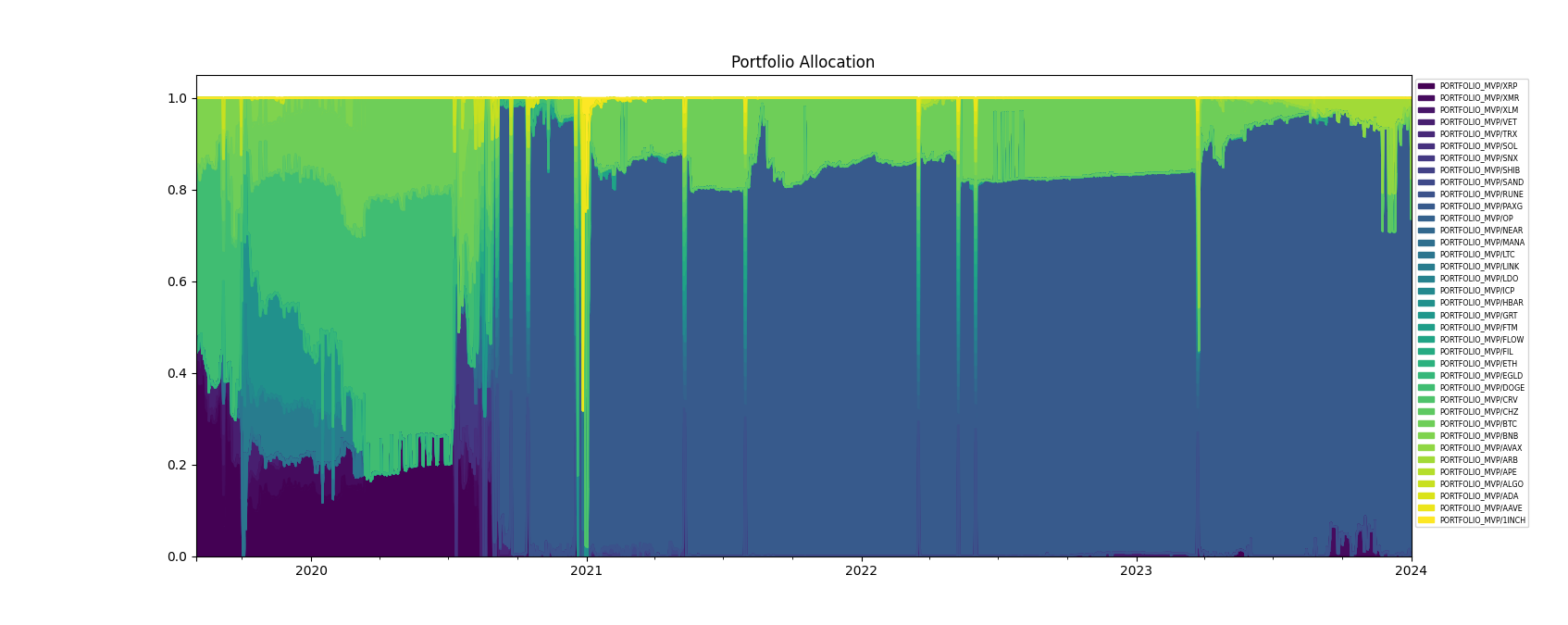}
        \caption{MVP Portfolio allocation (RW)}
        \label{fig:MVP_Weights_RW}
        \end{figure}

    \subsubsection{With Clustering}
        
        \begin{figure}[H]
            \centering
        \includegraphics[width=1.\columnwidth]{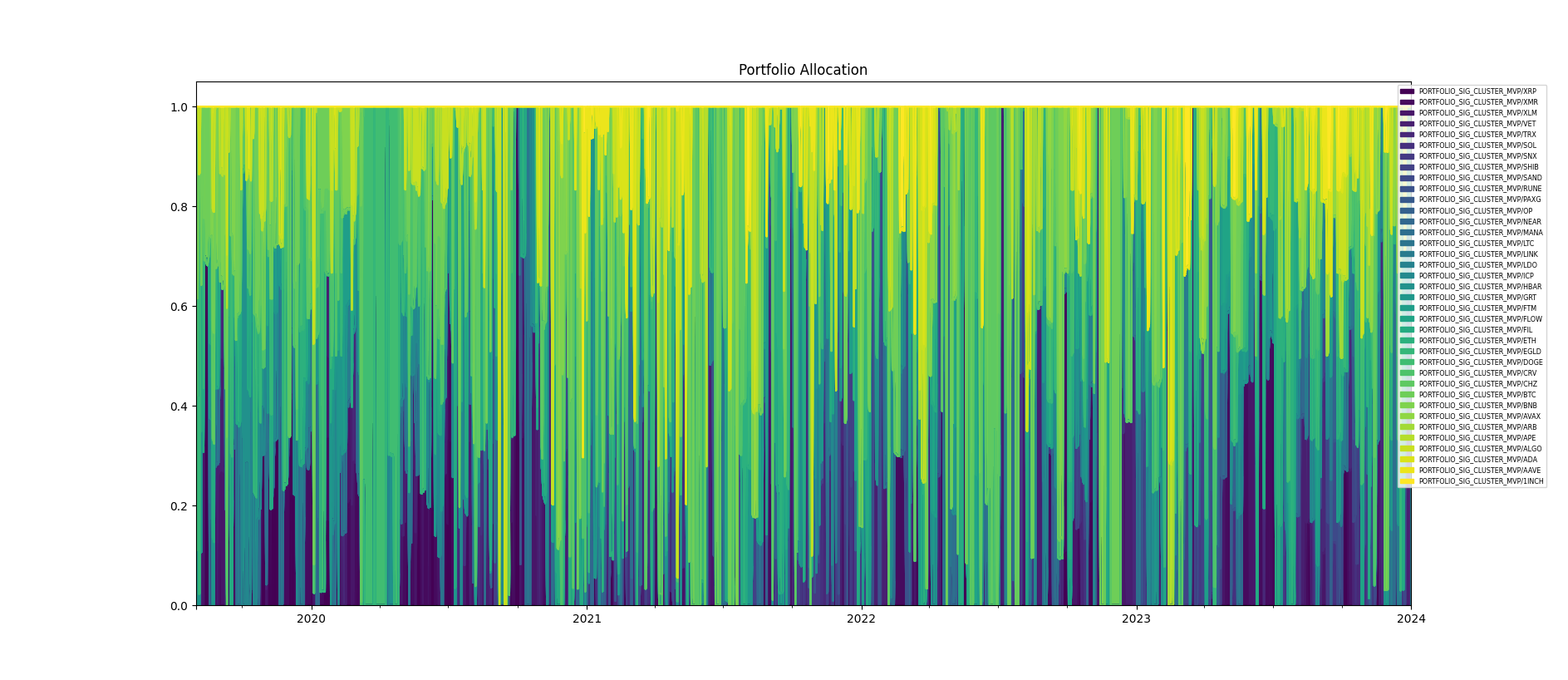}
        \caption{Signature Clustered MVP Portfolio allocation (RW)}
        \label{fig:MVP_Cluster_SIG_Weights_RW}
        \end{figure}

        \begin{figure}[H]
        \centering
        \includegraphics[width=1\columnwidth]{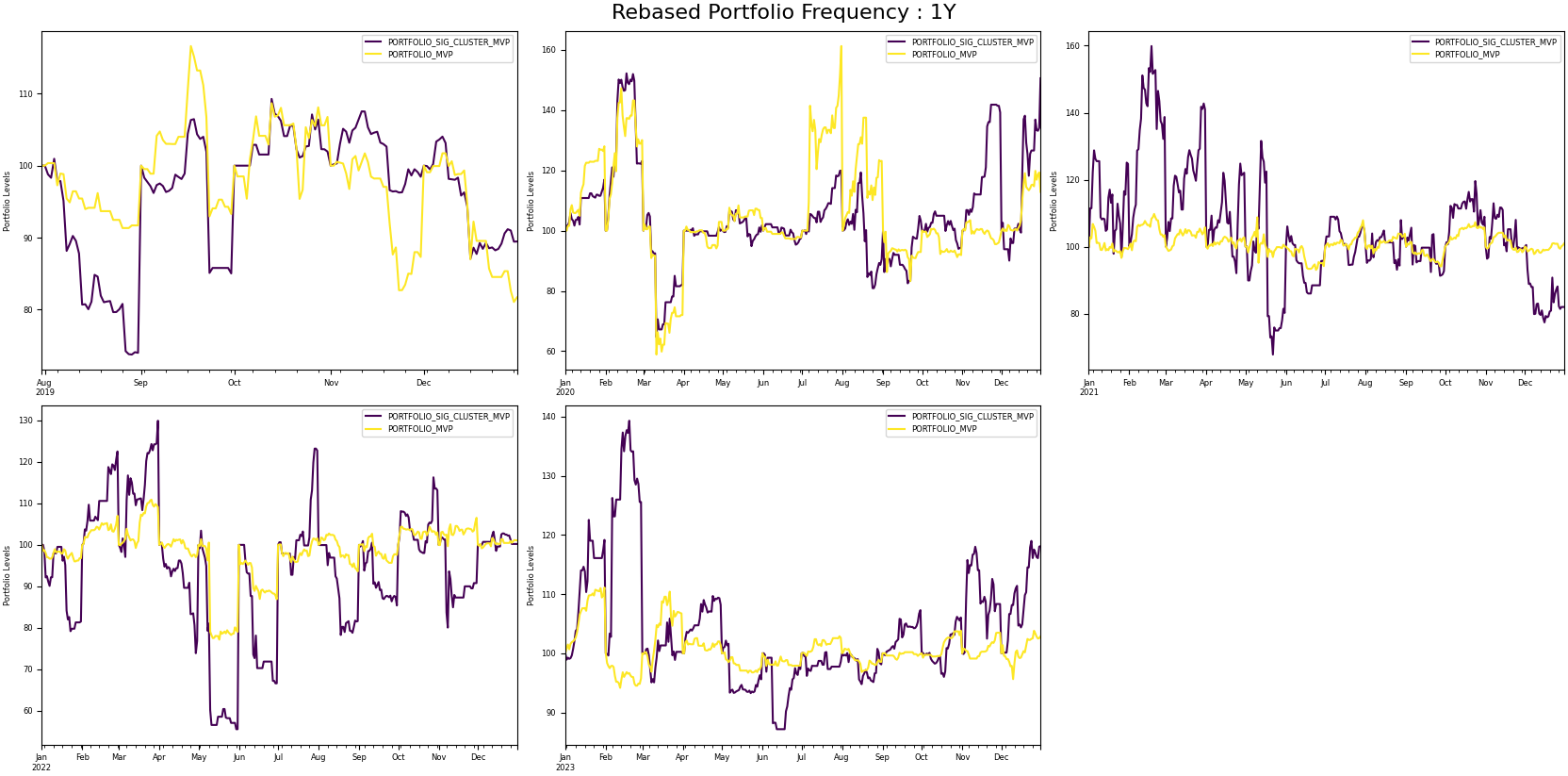}
        \caption{MVP Portfolio Rebased Annually (RW)}
        \label{fig:MVP_Rebased_Portfolio_RW}
        \end{figure}

\subsection{Maximum Diversification Portfolio (MDP)}

Table \ref{table:mdp_perf} shows the comparison of the annualized returns and annualized volatilities of two portfolios - the max diversification portfolio $(MDP)$ and the max diversification portfolio portfolio with clustering filter $(MDP_{sc}^{FOT})$ and $(MDP_{sc}^{RW})$. The $(MDP_{sc}^{FOT})$ shows a higher annualized return of 0.4013. However, it also has a higher level of annualized volatility at 0.6676 compared to 0.5742 for $(MDP)$. If we have a look at $(MVP_{sc}^{RW})$, it seems that the rolling window (RW) methodology has a really good impact to build the most diversified portoflio. We can clearly see the positive impact on the risk-return trade-off detailed Table \ref{table:mdp_risk}.

     \begin{table}[H]
        \centering
        \caption{Performance MDP Portfolios}
        \begin{tabular}{lcc}
        \toprule
            ~ & Annualized Return & Annualized Volatility  \\ \hline
            PORTFOLIO\_MDP & 0.2543 & 0.5742  \\ 
            PORTFOLIO\_SIG\_CLUSTER\_MDP\_FOT & 0.4013 &0.6676 \\ 
            PORTFOLIO\_SIG\_CLUSTER\_MDP\_RW & 1.1903 &0.7603 \\ 
        \bottomrule
        \end{tabular}
        \label{table:mdp_perf}
    \end{table}

    \begin{table}[H]
        \centering
        \caption{Risk metrics MDP Portfolios}
        \begin{tabular}{lccc}
        \toprule
            ~ & Sharpe & Calmar & MDD \\ \hline
            PORTFOLIO\_MDP & 0.4429 &0.3974 & 0.6394  \\ 
            PORTFOLIO\_SIG\_CLUSTER\_MDP\_FOT & 0.6011 & 0.5608 & 0.7151 \\ 
            PORTFOLIO\_SIG\_CLUSTER\_MDP\_RW & 1.5655 & 1.6362 & 0.7268 \\ 
        \bottomrule
        \end{tabular}
                \label{table:mdp_risk}
    \end{table}

  Table \ref{table:mdp_risk} shows the risk metrics for both portfolios, $(MDP_{sc}^{FOT})$ shows slightly higher Sharpe and Calmar ratios of 0.6011 and 0.4429, respectively. This indicates a better risk-adjusted return compared to $(MDP)$. Looking at the $(MDP_{sc}^{RW})$ we can clearly see the increase in the sharpe ratio using the RW methodology. There is a performance benefit to using the sliding window methodology, but the maximum drawdown is higher when clustering is used. It seems that the representative asset of the class picked during the portfolio universe construction is catching the most deterministic trend. Overall, both Table \ref{table:mdp_perf}-\ref{table:mdp_risk} suggest that $(MDP_{sc}^{RW})$ has a better risk-adjusted return than $(MDP)$ and $(MDP_{sc}^{FOT})$ portfolios. Despite this observation, it is important to note that $(MDP_{sc}^{RW})$ remains the most risky portfolio.

\subsection*{Fixed Origin of Time (FOT)}
 \textit{In this subsection you will find figures of the evolution of the portfolio allocation as well as the values of portfolios backtested with the methodology using the Fixed Origin of Time window (FOT)}
    \subsubsection{Without Clustering}

        \begin{figure}[H]
            \centering
        \includegraphics[width=1.\columnwidth]{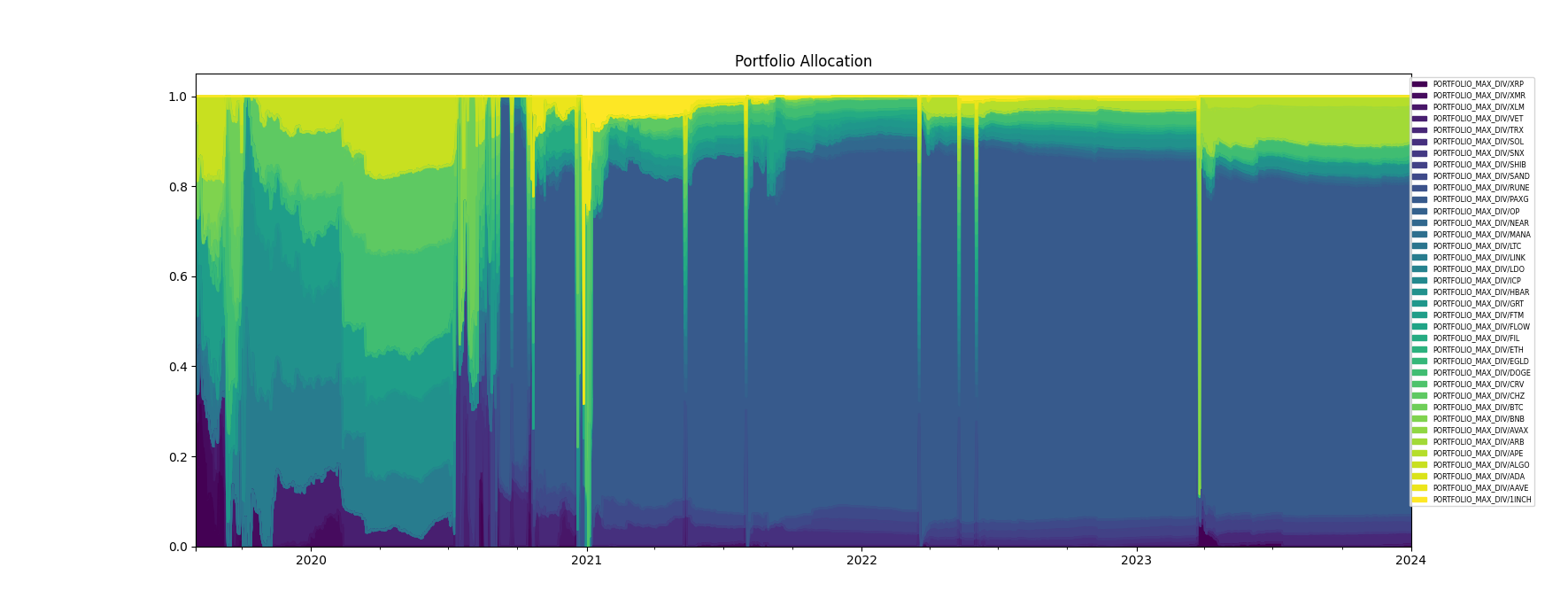}
        \caption{MDP Portfolio allocation (FOT)}
        \label{fig:MDP_weights}
        \end{figure}

    \subsubsection{With Clustering}
        
        \begin{figure}[H]
            \centering
        \includegraphics[width=1.\columnwidth]{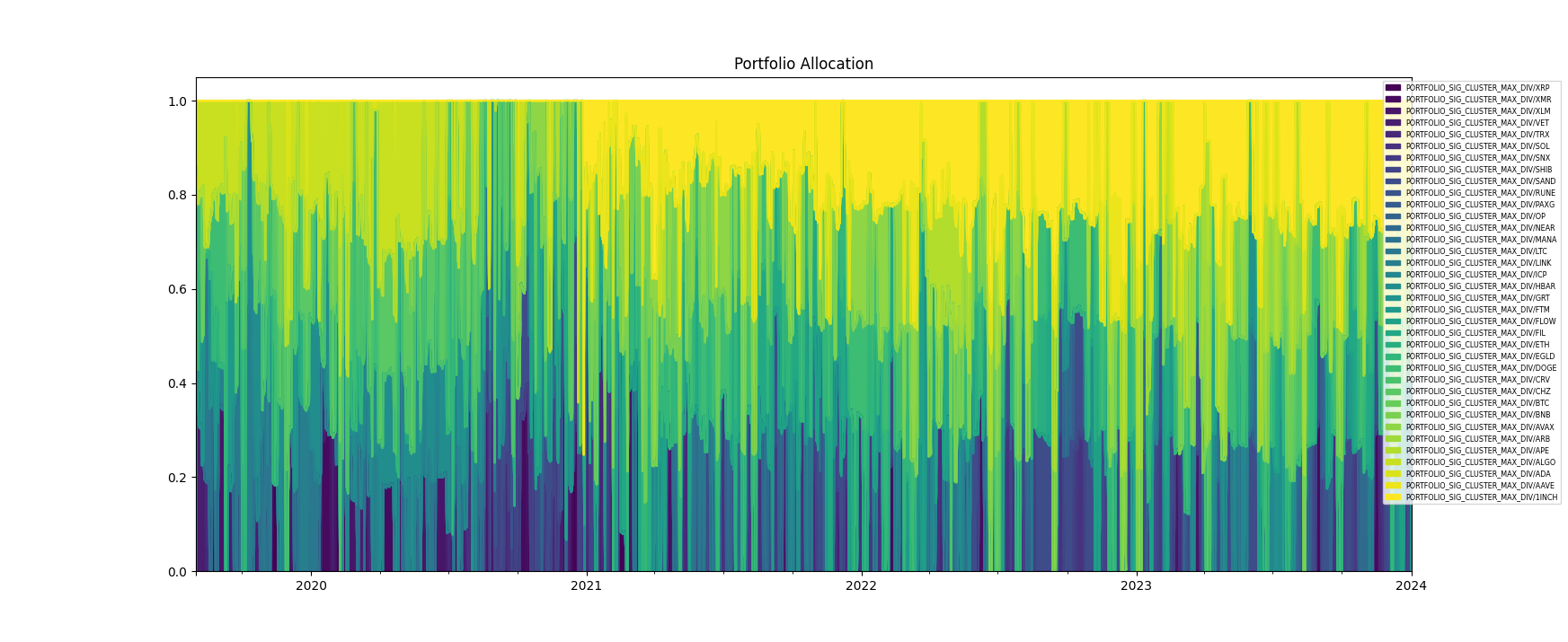}
        \caption{Signature Clustered MDP Portfolio allocation (FOT)}
        \label{fig:MVP_Cluster_SIG_Weights}
        \end{figure}
    
        \begin{figure}[H]
        \centering
        \includegraphics[width=1\columnwidth]{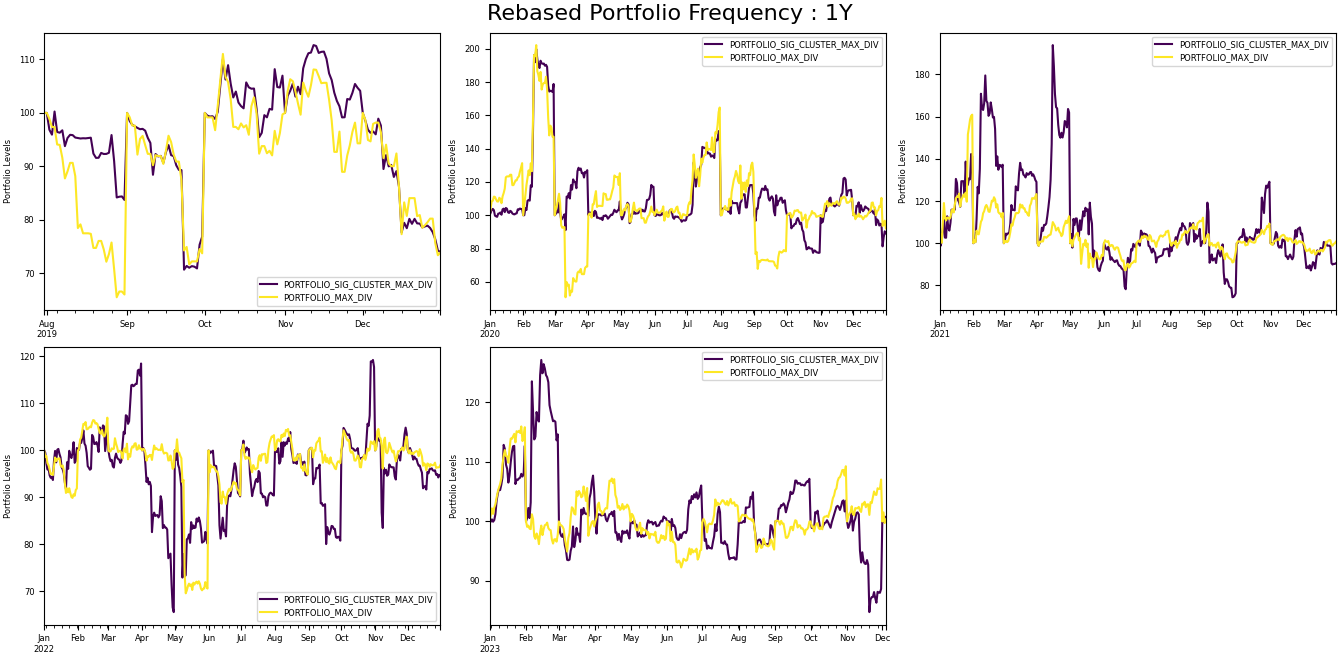}
        \caption{MDP Portfolio Rebased Annually (FOT)}
        \label{fig:MDP_Rebased_Portfolio}
        \end{figure}
\subsection*{Rolling Window (RW)}
    \textit{In this subsection you will find additional figures of the evolution of the portfolio allocation as well as the values of portfolios backtested with the methodology using the rolling window (RW)}
  \subsubsection{Without Clustering}

        \begin{figure}[H]
            \centering
        \includegraphics[width=1.\columnwidth]{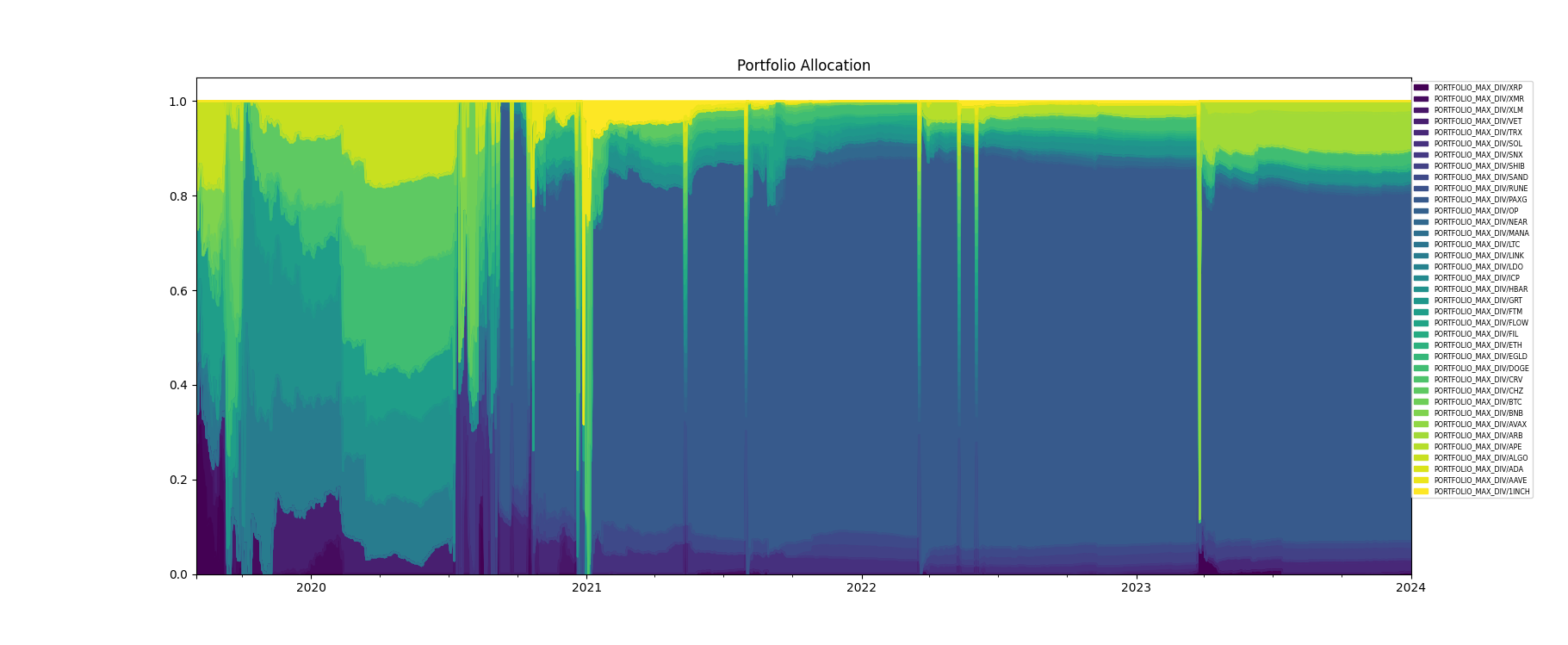}
        \caption{MDP Portfolio allocation (RW)}
        \label{fig:MDP_Weights_RW}
        \end{figure}

    \subsubsection{With Clustering}
        
        \begin{figure}[H]
            \centering
        \includegraphics[width=1.\columnwidth]{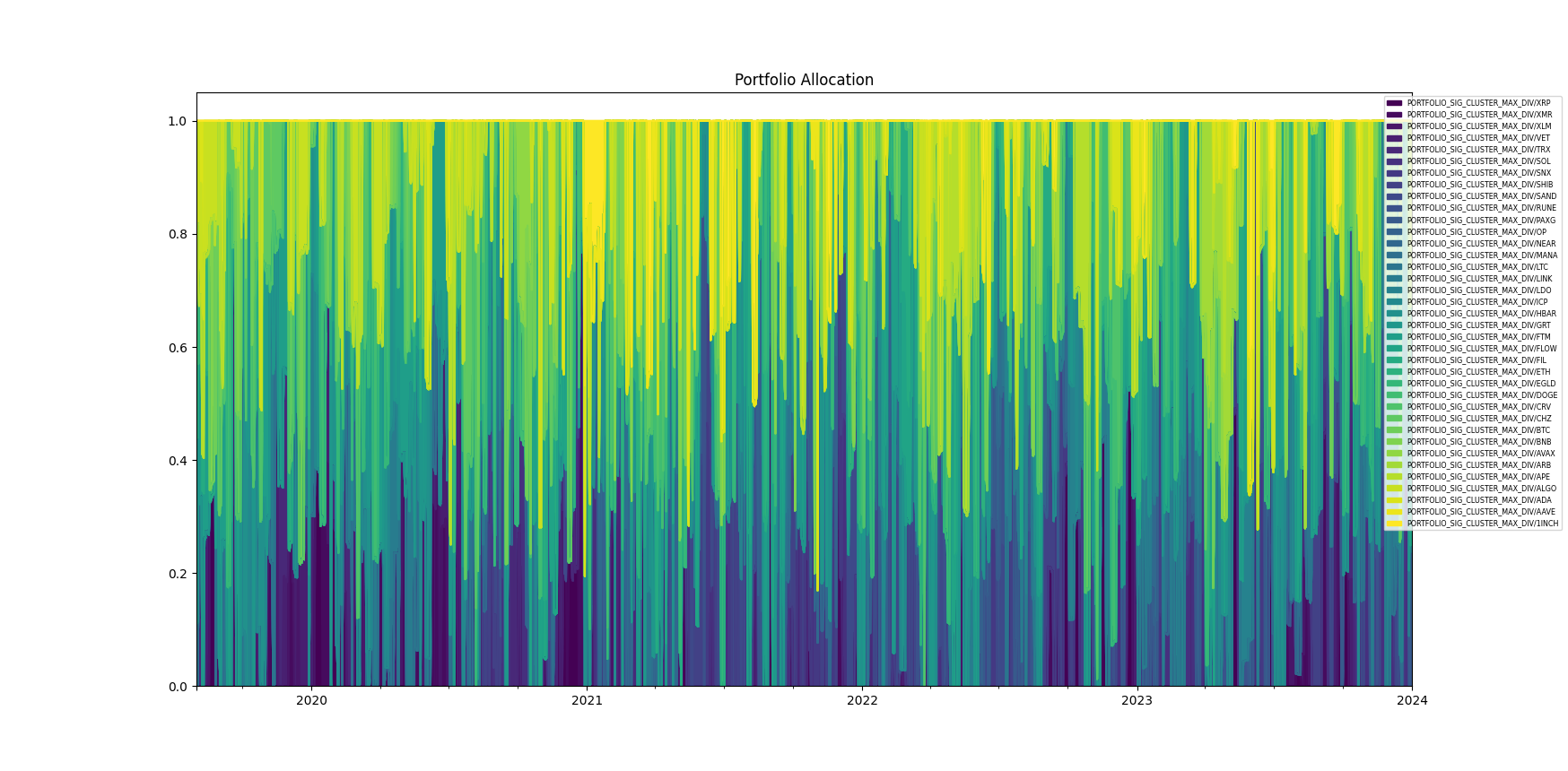}
        \caption{Signature Clustered MVP Portfolio allocation (RW)}
        \label{fig:MDP_Cluster_SIG_Weights_RW}
        \end{figure}

        \begin{figure}[H]
        \centering
        \includegraphics[width=1\columnwidth]{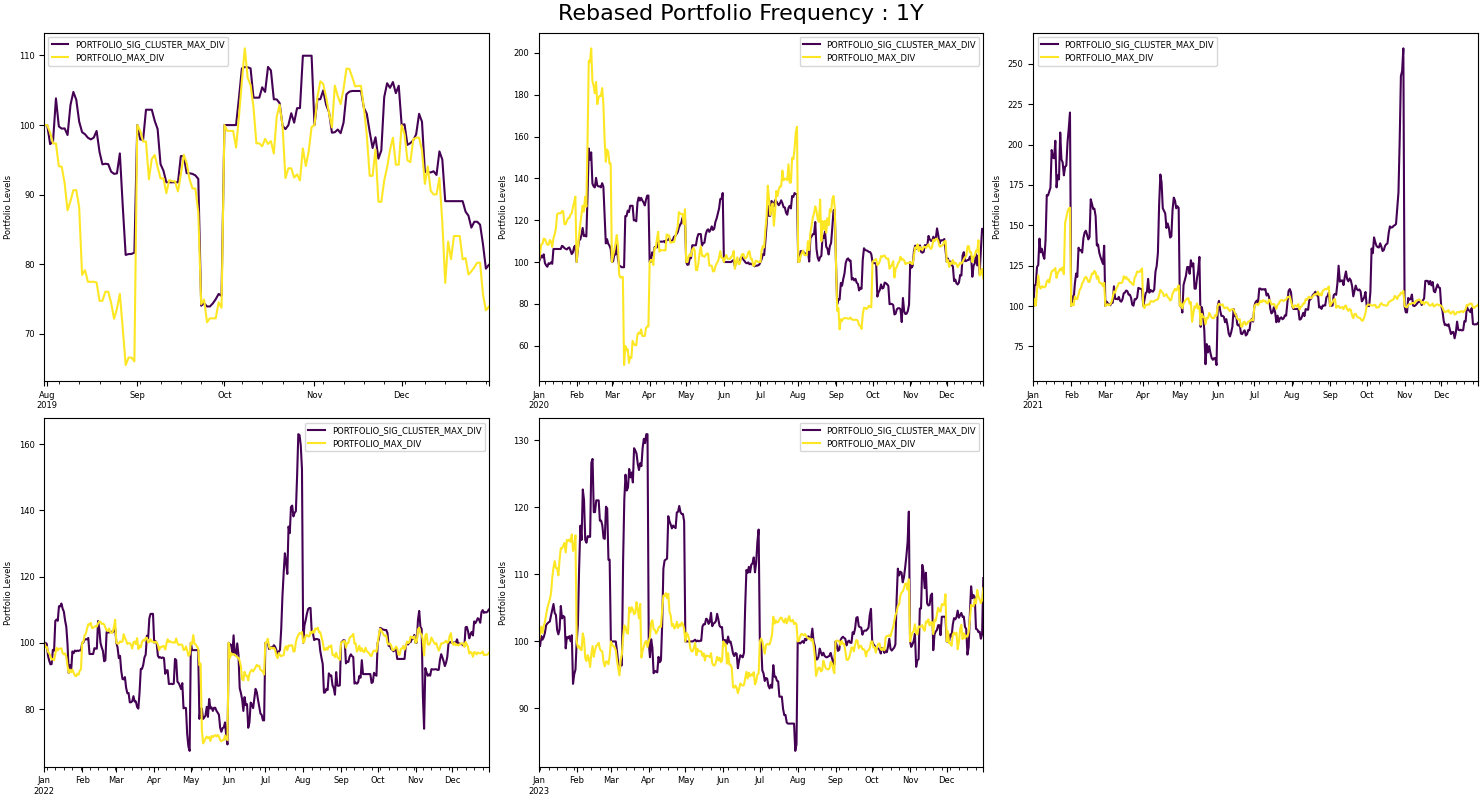}
        \caption{MDP Portfolio Rebased Annually (RW)}
        \label{fig:MDP_Rebased_Portfolio_RW}
        \end{figure}

\subsection{Comparison}

\begin{figure}[H]
\centering
\begin{minipage}{.5\textwidth}
  \centering
  \includegraphics[width=.8\linewidth]{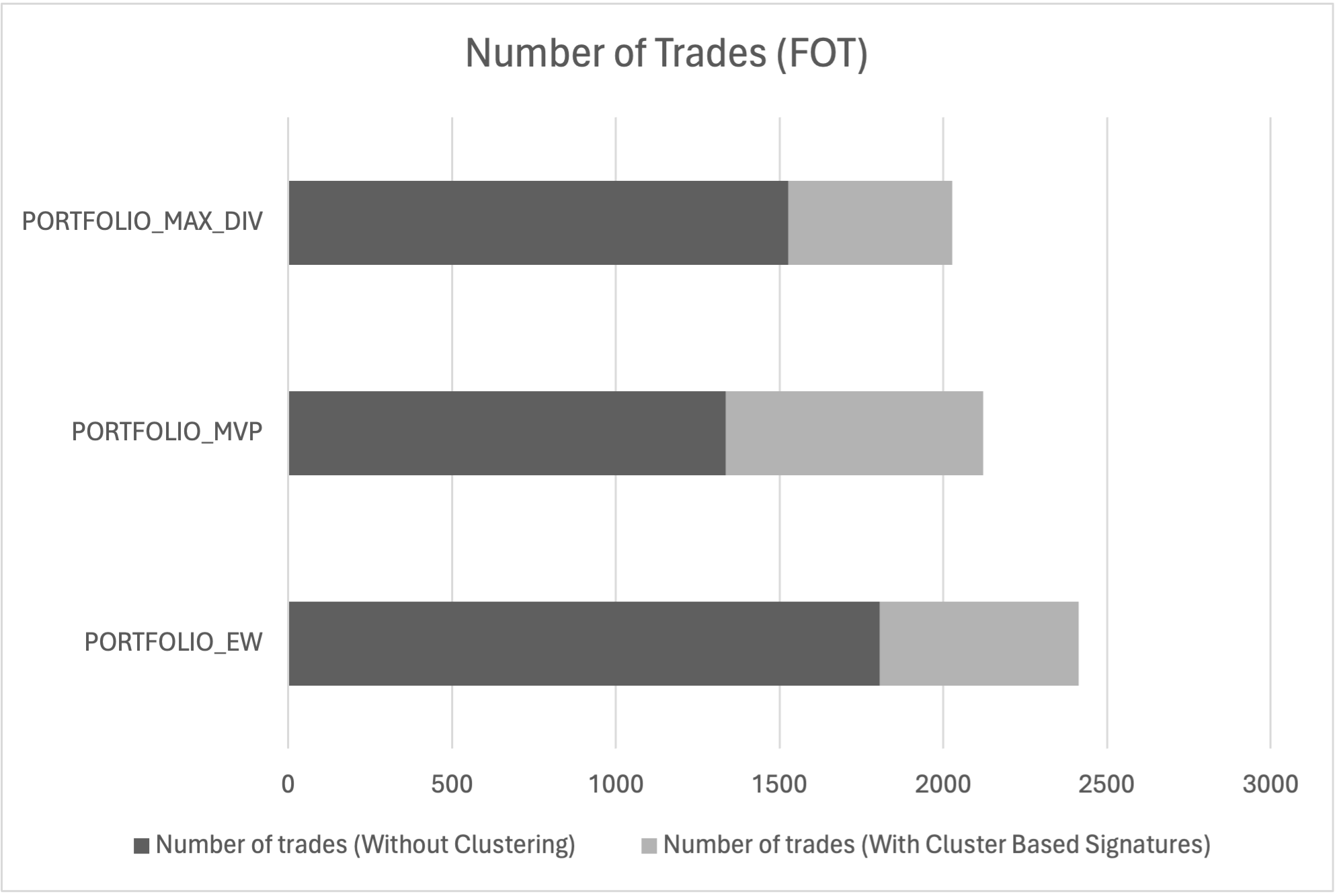}
  \captionof{figure}{Number of trades (FOT)}
  \label{fig:nb_of_trades_FOT}
\end{minipage}%
\begin{minipage}{.5\textwidth}
  \centering
  \includegraphics[width=.8\linewidth]{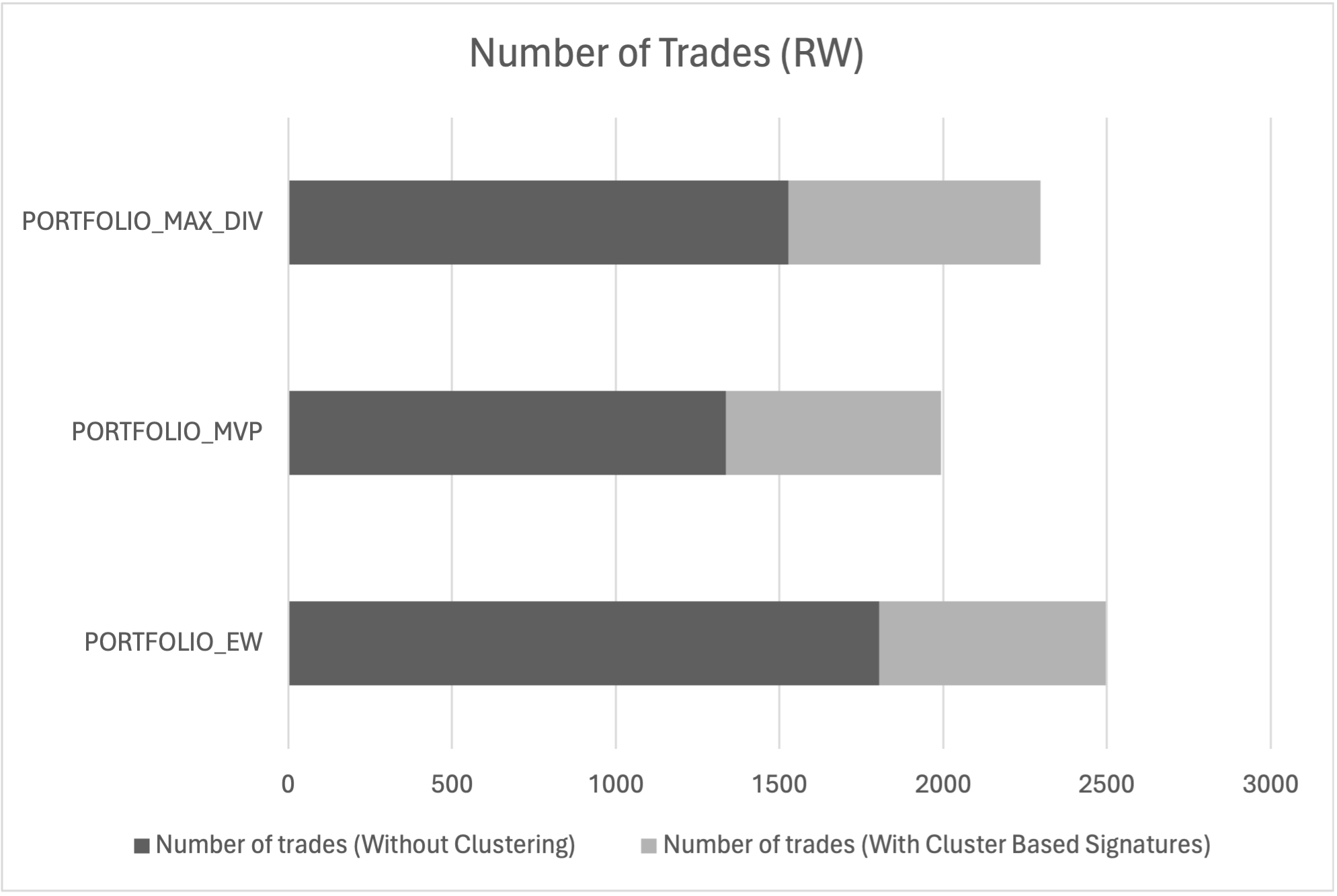}
  \captionof{figure}{Number of trades (RW)}
  \label{fig:nb_of_trades_RW}
\end{minipage}
\end{figure}

Figures \ref{fig:nb_of_trades_FOT}-\ref{fig:nb_of_trades_RW} shows the number of trades for each methodology, using the Fixed of Time (FOT) window and the rolling window (RW). On both figures we can clearly identify the less expensive strategies in term of trading cost. For our backtest we used trading fees of 20 bp. Considering that asset managers may play these strategies we did not considerered a fixed trading fees for each trading, primarily because of the unrealism it suggests.

\section{Conclusion}

In this chapter, we have introduced a new method to classify financial assets, which differs from traditional approaches centered on historical returns. We have chosen to exploit path signatures, offering a robust alternative for synthesizing the information contained in digital asset prices. This method enables us to discover a different representation, facilitating the identification of similarities between various assets. It also allows us to operate in a higher-dimensional space, while taking time dependency into account. The results obtained with different portfolios clearly demonstrate the advantages of this approach, particularly for investors. For the $(EW)$ portfolio, the clustering-filtered version denoted $(EW_{sc}^{FOT})$ and $(EW_{sc}^{RW})$ significantly outperforms the standard (unfiltered) methodology in terms of  annualized returns (0.9592, 1.2523 vs. 0.5984) and  annualized volatility (0.6319, 0.7432 vs. 0.8219). This superior performance is further supported by higher Sharpe and Calmar ratios, and a lower Maximum DrawDown (MDD), indicating a more efficient risk-adjusted return and a lesser decline in value over time. A similar analysis with Mean-Variance portfolios shows that the clustering-filtered MVP $(MVP_{sc}^{FOT})$ also has a higher annualized return (0.2199 vs. 0.1140) compared to the standard$(MVP)$. However, $(MVP_{sc}^{RW})$ reports a weak performance of 0.1488 for a level of risk equivalent to the $(MVP_{sc}^{FOT})$. If we have a look on risk metrics there is a slightly better risk-adjusted returns for $(MVP_{sc}^{FOT})$, but tempered by a higher MDD. Finally, the $(MDP)$ comparison reveals that the clustering-filtered MDP denoted $(MDP_{sc}^{FOT})$ and $(MDP_{sc}^{RW})$ yield a higher annualized return (1.1903, 0.4013 vs. 0.2543) than the standard MDP. The cost of the higher return is the risk of the portfolio, clustering-filtered versions shows an higher annualized volatility (0.7603, 0.6676 vs. 0.5742).  However, the risk metrics indicate better risk-adjusted returns for $(MDP_{sc}^{RW})$, although it experiences a higher MDD.  Overall, the clustering-filtered portfolios consistently demonstrate higher returns compared to their standard counterparts, except for the $(MVP)$. However, these benefits are accompanied by increased volatility and MDD, indicating a trade-off between higher returns and increasing risks. Investors should weight these factors carefully, considering their risk tolerance and investment objectives.

\bibliography{bib.bib}

\end{document}